\documentclass[bachelor]{iisthesis}

\pdfoutput=1

\usepackage[T1]{fontenc} 
\usepackage{lmodern}     
\usepackage{amsthm}      
\usepackage{multirow}
\usepackage{amsfonts}
\usepackage{amsmath}
\usepackage{graphics}    
\usepackage{subfigure}
\usepackage{tikz}
\usepackage{listings}
\usetikzlibrary{shapes, arrows, calc, fit, positioning, decorations.markings}

\title{Automatic Determination of Chord Roots}
\author{Samuel Rupprechter}
\supervisor{Univ.-Prof.~Dr.~Justus Piater}

\begin{document}

\theoremstyle{plain}
\newtheorem{corollary}{Corollary}[%
]
\newtheorem{hypothesis}[corollary]{Hypothesis}
\theoremstyle{definition}
\newtheorem{definition}[corollary]{Definition}

\newcommand{\lilypondinclude}[1] {
    \begin{quote}
    {%
        \parindent 0pt
            \noindent
            \ifx\preLilyPondExample \undefined
            \else
            \expandafter\preLilyPondExample
            \fi
            \def\lilypondbook{}%
            \begin{center}
            \includegraphics{#1}
            \end{center}
        \ifx\postLilyPondExample \undefined
            \else
            \expandafter\postLilyPondExample
            \fi
    }
    \end{quote}
} 

\newcommand\wideDownarrow{\mathrel{\scalebox{2.8}[2.5]{$\downarrow$}}}

\maketitle

\chapternn{Abstract}

Even though chord roots constitute a fundamental concept in music theory,
existing models do not explain and determine them to full satisfaction.
We present a new method which takes sequential
context into account to resolve ambiguities and detect nonharmonic tones. We
extract features from chord pairs and use a decision tree to determine chord
roots. This leads to a quantitative improvement in correctness of the predicted
roots in comparison to other models.
All this raises the question how much harmonic and nonharmonic tones actually
contribute to the perception of chord roots.

\tableofcontents
\listoffigures
\listoftables
\uibkdeclaration{}
\label{chap:declare}

\chapter{Introduction}

\section{Motivation}

Chord roots are of central importance in many theories of music. Therefore, it
is perhaps surprising that only little research has been done in that area and
many characteristics of chord roots are unknown.  Several already existing
theories for the prediction of chord roots have been shown to fail
by~\cite{Goldbach2009}. Martin Anton Schmid, a Tyrolean music theorist, formed a
small research team with the goal to develop a new model to determine and
explain chord roots.  One of the new ideas from this group is to use the
harmonic context of chords.


In this thesis we present the new basic model and an extension using the just
mentioned context of chords.
As far as we know, this is the first take on the topic of chord roots from a
computer science perspective.

The model we describe is by no means finished nor do we claim it will work for
every kind of music, but it is a first step towards a better understanding of
chord roots in classical music. A long term goal of the determination of roots
in musical pieces would be a harmonic analysis of chord progressions for
various musical genres.\footnote{This could one day even include
non-Western tonalities.}

\section{Goals}

Since this project entails basic research, it was not always clear how far we
would come during the course of this thesis. Nevertheless, the goals of this
project were to extend the Schmid model (which we will introduce in
Section~\ref{chap:SchmidModel}) with musical context and to write a
program that
\begin{itemize}
    \item reads in existing pieces of music\footnote{For this thesis we focused
    on pieces from Johann Sebastian Bach. The BWV numbers abbreviate
``Bach-Werke-Verzeichnis'' (Bach Works Catalogue) and all the BWV-examples we use in
this thesis are taken from the music21 corpus~\citep{music21}.} from some machine-readable format,
    \item transforms these pieces into a newly developed representation,
    \item determines the roots of the chords in these pieces using the new model as well as other existing models.
\end{itemize}

\noindent All these goals have been fulfilled and in addition to that 26 pieces
of music (containing 2290 chords\footnote{Although maybe not chords in the
traditional sense. Cf.~Section~\ref{chap:Chordification}}) have been annotated
by hand with the ``correct roots''\footnote{These annotated roots correspond to
my own subjective intuition of what the correctly predicted roots should be when
taking nonharmonic tones into account and are most likely controversial and
might contain errors.} for each chord. Using musical context we were able to
determine 95.34\% of the roots
correctly.\footnote{Appendix~\ref{chap:AppendixBenchmarks} contains benchmarks
for all models.}

Note that we focused on the development of a new algorithm and no empirical
studies have been done to confirm or deny the results. This will have to be one
of the next steps of our research group.

\section{Overview}

In Chapter~\ref{chap:MusicTheory} we will give a quick introduction to basic
music theory, which is a prerequisite to understand the following chapters.
Chapter~\ref{chap:DeterminationChordRoots} focuses on a specific part of this
theory, namely the determination of chord roots. It describes several existing
models and then gives an explanation of the Schmid model.
Chapter~\ref{chap:MusicalContext} then introduces an extension of this model
using musical context of chords, which is classified via a decision tree that we
created manually, while Chapter~\ref{chap:AutomaticDecisionTree} explains how we
created a different decision tree automatically.
Chapter~\ref{chap:Implementation} gives an overview of the implementation and
explains how to use the program.


\chapter{Music Theory}
\label{chap:MusicTheory}

This chapter will give a short but for our purposes sufficient introduction to
basic aspects of music theory. For a more thorough description please refer
to~\cite{Ziegenruecker2009},~\cite{Grabner1974} or~\cite{Schoenberg2010}.  Even
if you are familiar with music theory you might want to quickly read through the
definitions of chords and chord roots in this chapter, because we also explain
the simplifying assumptions that we used.

\section{Basic Music Theory}

\subsection{Musical Notation}

\begin{figure}[h]
    \lilypondinclude{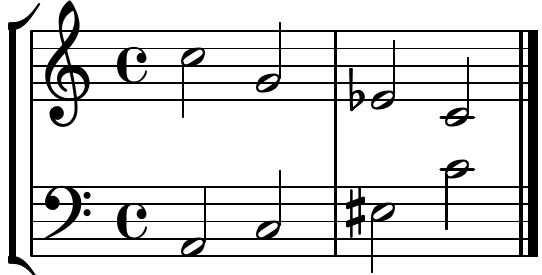}
    \caption{An example of basic musical notation.}
    \medskip
    \small
    \begin{center}
    The two staffs have different clefs at the beginning, which makes the last
    note in both staffs a \emph{Middle~C} ($C_4$ in Scientific pitch notation or
    $c'$ in Helmholtz notation), which has a frequency of
    \emph{261.626 Hertz} (in equal temperament with standard pitch $A_4 = 440 \;
    Hertz$). The second note in the second staff is a \emph{Bass~C}
    ($C_3$ or $c$) with exactly half the frequency of the Middle~C. We say it is an
    \emph{octave} below the $C_4$.
    \end{center}
\label{fig:MusicalNotationExample}
\end{figure}

To represent a piece of music we essentially use a graph of \emph{pitch} versus
\emph{time}. The pitch of a music note is essentially the base frequency of that
note---and of course higher frequencies make higher sound. We call the
horizontal lines a \emph{staff} and the symbol at the beginning of the staff the
\emph{clef}.
The vertical lines represent measures, which we will ignore in this
introduction.
Clefs, the symbols at the beginning, indicate what pitches the written notes have
on that staff. Several staves can be written above each other, which means they
are intended to be played at the same time (either by different instruments or,
as in the case of the piano, by different hands).
Figure~\ref{fig:MusicalNotationExample} shows an example of all of these
elements and more.

\subsection{Notes and Pitches}

The dots written on a staff are called \emph{notes}. The pitches of these notes
can be altered using \emph{accidentals} (in addition to being determined by the
clef). The only two accidentals we will use in this thesis are the \emph{sharp}
($\sharp$), which raises a note by a half step and the \emph{flat} ($\flat$),
which lowers a note by a half step. A \emph{half step} is the distance two
adjacent keys on the piano keyboard, regardless of what color they may have.  A
\emph{step} is equal to two half steps.  Note that this means that it is
possible to represent the same pitch in multiple ways using different notes and
accidentals (e.g. $C$ and $B\sharp$). We call this \emph{enharmonic
equivalence}\footnote{This is only possible in the equal temperament, which we
assume here.}.

\subsubsection{Note Names}

The white keys on the piano keyboard are given names according to the English
alphabet from $A$ to $G$---and then starting with $A$ again. This next $A$ will
have exactly double the frequency as the first $A$ and we say it is an
\emph{octave} higher. In order to distinguish between notes with the same name,
the name (or number) of the octave can added to the name of the note.

\subsubsection{Pitch Classes}

An important notion for this thesis is that of \emph{pitch classes}. It is a
simple mapping of note names (starting from $C$) to natural numbers between $0$
and $11$~\citep{forte1973}. Octaves are ignored and the mapping is irreversible (without
ambiguity).  Table~\ref{tab:PitchClassesExample} shows notes and
their corresponding pitch classes in integer notation with $C$ as $0$.

\begin{table}[h]
\begin{center}
\begin{tabular}{|c c|c|}
    \hline
    Note Name & Alternative Name & Pitch Class \\
    \hline
    $C$       & $B\sharp$ & 0  \\
    $C\sharp$ & $D\flat$  & 1  \\
    $D$       &           & 2  \\
    $D\sharp$ & $E\flat$  & 3  \\
    $E$       & $F\flat$  & 4  \\
    $F$       & $E\sharp$ & 5  \\
    $F\sharp$ & $G\flat$  & 6  \\
    $G$       &           & 7  \\
    $G\sharp$ & $A\flat$  & 8  \\
    $A$       &           & 9 \\
    $A\sharp$ & $B\flat$  & 10 \\
    $B$       & $C\flat$  & 11 \\
    \hline
\end{tabular}
\caption{Note names (without octaves) and their corresponding pitch classes.}
\label{tab:PitchClassesExample}
\end{center}
\end{table}

\subsection{Rhythm}

\begin{figure}
    \lilypondinclude{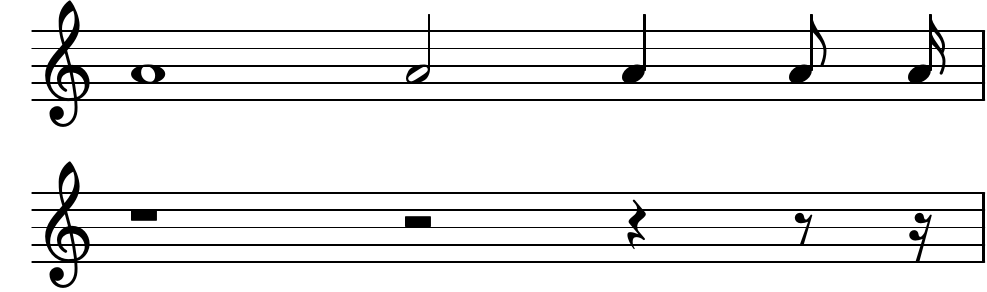}
    \caption{Common lengths of notes and rests.}
    \medskip
    \small
    \begin{center}
    The first note is a \emph{whole note} and each following note has half the
    length of the \\ previous one. The same applies to the rests in the second
    staff.
    \end{center}
\label{fig:RhythmExample}
\end{figure}

Notes of course do not only have a pitch, but also a \emph{length}, which is
what gives a piece its \emph{rhythm}. A special notation for ``a time of
silence'' also exists and is called a \emph{rest}. Rests can also have different
lengths. Note that these lengths do not say anything about the absolute time a
note should be played (or nothing should be played in the case of the rest), but
rather it notates the time it should take in relation to the other notes.
Figure~\ref{fig:RhythmExample} shows an example of both notes and rests with
common lengths.

\subsection{Intervals}

\begin{figure}[h]
    \lilypondinclude{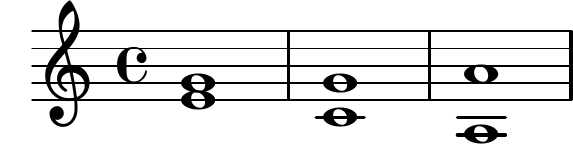}
    \caption{The three intervals \emph{Minor third}, \emph{Perfect fifth} and
    \emph{Perfect octave}.}
\label{fig:IntervalsExample}
\end{figure}

The last important notion in this section is that of the \emph{interval}, which
describes the distance between two pitches. To calculate this distance, simply
count the number of half steps it takes to reach the top note from the bottom
one. These distances are then again given names\footnote{Actually it is
necessary to count both steps and half steps, because
enharmonics have the same number of half steps but different intervals.
See Table~\ref{tab:IntervalsComplete} in
Appendix~\ref{chap:Intervals} for a more detailed list of intervals.}.
Figure~\ref{fig:IntervalsExample} shows some examples of intervals that will be
needed later on.

\section{Chords}

We decided to take the definition from~Schmid~(2014, personal communication):

\begin{definition}[Chord]

    A chord is a combination of two or more tones with different pitch classes
    sounding simultaneously or immediately in succession.

\end{definition}

Schmid also claims that ``The connection between between successive tones
can sometimes only be determined in the harmonic context and sometimes not at
all.''

\begin{figure}[h]
    \begin{center}
    \subfigure[A single note]{
        \includegraphics{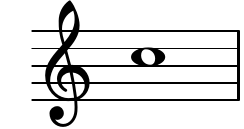}
        \label{fig:ValidInvalidChordExample:subfig1}
    }
    \subfigure[A broken $Cmaj$ chord]{
        \includegraphics{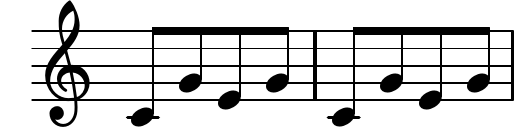}
        \label{fig:ValidInvalidChordExample:subfig2}
    }
    \subfigure[An interval]{
        \includegraphics{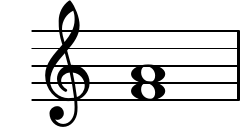}
        \label{fig:ValidInvalidChordExample:subfig3}
    }
    \subfigure[$A\flat{}maj7$ chord]{
        \includegraphics{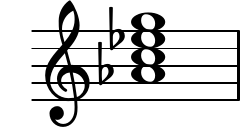}
        \label{fig:ValidInvalidChordExample:subfig4}
    }
    \end{center}
    \caption{A single note and three note combinations, of which we only
    consider (c) and (d) as ``chords''.}
\label{fig:ValidInvalidChordExample}
\end{figure}

With this definition we do not distinguish between \emph{intervals} (i.e.~chords
with only two notes) and chords with three or more notes.
In addition to that
we make another simplification: We do not view arpeggios, broken chords and so
forth (i.e.~the tones sounding immediately in succession) as chords, because
simply not enough research has been done in this area. It is for example not
clear how fast (in absolute time) a broken chord has to be played to be
perceived as a chord (with a root).
It will probably be necessary to revise this definition after further research.
Figure~\ref{fig:ValidInvalidChordExample} shows several examples of note
combinations. Only the third and fourth example are chords in the sense of our
simplified definition.

\begin{definition}[Chord Progression]
    A chord progression is a sequence of two or more chords.
\end{definition}

A chord progression usually gives a piece of music its \emph{harmonic movement}
and is sometimes even called \emph{harmonic progression}~\citep{Schoenberg1969}.
It is the main reason we want to determine chord roots, which we will define
shortly.

\subsection{Nonharmonic Tones}

\begin{definition}[Nonharmonic Tone]
    A nonharmonic tone (also called nonchord tone) is a note in a chord which
    does not fit into the established harmonic framework. A chord may contain
    zero, one or more nonharmonic tones.
\end{definition}

Nonharmonic tones are almost always dissonant (Schmid, 2015, personal
communication), which
means the chord they are played in will sound ``harsh'' or ``unpleasant'', in
contrast to a chord containing only consonant tones, which will sound ``sweet''
or ``pleasant''. Note that, as~\cite{Hindemith1942} stated, the two concepts of
consonance and dissonance ``have never been completely explained, and for a
thousand years the definitions have varied''.

\begin{figure}[h]
    \lilypondinclude{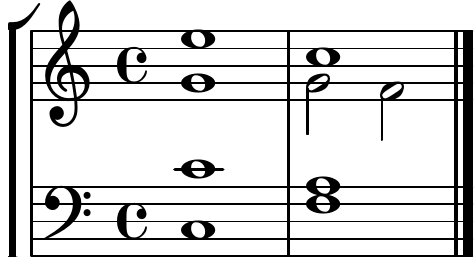}
    \caption{Example of a suspension ($G$) in the second measure.}
\label{fig:SuspensionExample}
\end{figure}

Music theorists distinguish between many different forms of nonharmonic tones,
which are described in most introductions to music theory.
If we know
the root of a chord, its nonchord tones can in most cases easily be determined,
although we did not focus on that in this thesis.
Figure~\ref{fig:SuspensionExample} shows an example of a \emph{suspension}---a
common type of nonharmonic tones in the Bach corpus.

\section{Chord Roots}


We used the following working definition:

\begin{definition}[Chord Root]
    The root of a chord is its perceptional \emph{reference note}, on which the chord
    progression of a piece can be analysed.
    A chord might have more than one correct root.
    Sometimes the root(s) of a chord cannot be determined.
\end{definition}

\begin{figure}[t]
    \lilypondinclude{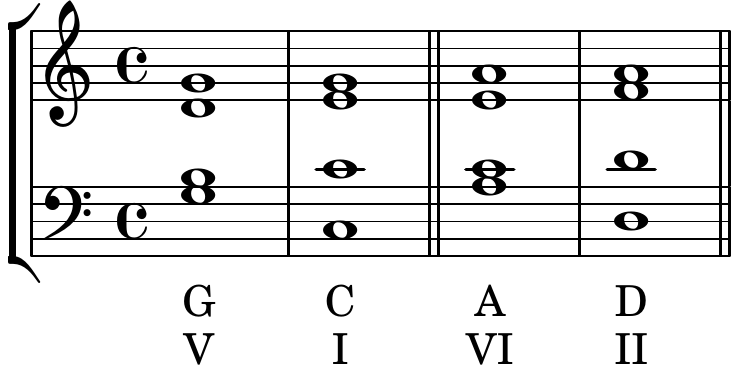}
    \caption{An example of the same chord progression created by different chords.
    \citep{Schmid2014}}
\label{fig:ChordRootsIntroductionExample}
\end{figure}

Figure~\ref{fig:ChordRootsIntroductionExample} shows an example of a simple
chord progression with the roots of the chords and roman numerals written below
each chord\footnote{We will see how these roots can be determined in the next
chapter.}.  $G$ is 4 steps above $C$ and $A$ is 4 steps above $D$, which means
the chord progressions are actually the same: ``four steps down''.

\chapter{Determination of Chord Roots}
\label{chap:DeterminationChordRoots}

In this chapter we describe how chords can be created by stacking thirds and
how several existing models for the determination of roots analyse chords.
Then Schmid's new model is introduced. Finally, it is shown how chords are
extracted from note sequences and an analysis example of such a sequence is
given.

\section{Constructing Chords}
\label{chap:GeneratingChords}

The classical way to generate a chord would be to take a note as root (e.g. $C$)
and stack a third and a fifth above it. For a $Cmaj$ chord you would take a
Major third ($E$) and a Perfect fifth ($G$) to get $C-E-G$. For $Cm$ the Minor
third is needed first and then again the Perfect fifth ($C-E\flat-G$).  Note
that in both cases the highest note is a third above the second highest note: A
Minor third in the first and a Major third in the second chord. If we stack
another third above each chord we get a $Cmaj7$ ($C-E-G-B$) or $Cm7$
($C-E\flat-G-B\flat$) chord.
This method of generating chords goes back to~\cite{Rameau1722}.

\section{Existing Models}

As mentioned in the beginning,~\cite{Goldbach2009} showed that all existing
models fall short of correctly determining chord roots in many cases.
Nevertheless, they are quite interesting and useful for comparisons for the
model of Schmid and we will have a look at them in the remainder of this section.
We will do a detailed analysis of the chord $E_4-G_4-C_5$\footnote{A $Cmaj$
chord in its first \emph{inversion}.} by each model after
describing it. These models were also implemented in our program.


\subsection{Stacking Thirds}

\begin{figure}[h]
    \begin{center}
        \lilypondinclude{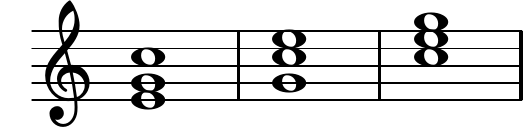}
    \caption{Analysis of a $Cmaj$ chord in its first inversion by stacking thirds.}
    \medskip
    \small
        $E$ has one third above it, $G$ has zero thirds directly above.  $C$ can
        have both the other notes stacked above it and we determine it as the
        root.
    \end{center}
\label{fig:CmajStackingThirds}
\end{figure}

This describes most well known method of finding the root of a chord (actually
it is the method that is taught in most music classes to determine the
\emph{name} of a chord, which does not necessarily coincide with its
\emph{perceptual} root).
We basically try to reverse the process of generating a chord that we just
introduced in Section~\ref{chap:GeneratingChords}:

Ignore octaves and stack as many notes as possible with distance of a third
above each other. The lowest note will be the root. Note that this addresses
primarily how a chord is notated in contrast to how a chord sounds, which is
what the following models will do. Figure~\ref{fig:CmajStackingThirds} shows an
example of this method.

\subsection{Ernst Terhardt}

\cite{Terhardt1982} based his model on the first ten \emph{subharmonic partials} of each note
in a chord:

\begin{definition}[Subharmonic Partial]
    The n'th subharmonic partial P of a note $X$ with frequency $f_X$ is a tone with
    frequency $f_P = f_X/n \;$ with $ \; n \in \mathbb{N} \setminus \{0\}$.
\end{definition}

Ignoring octaves, each note in a chord has five different notes in its first ten
subharmonic partials (which include the note itself). These partials correspond
(approximately) to the following intervals: Perfect unison, Perfect fifth below,
Major third below, Major second above and Major second below.  After having
determined the subharmonic partials for each note in the chord, the frequency of
each note is counted. Each note that appears $d$ times in the subharmonic
partials of the notes in the chord is determined as a ``root of degree $d$'' of
the chord.

The note with the highest degree (which is not necessarily unique) is then
determined as ``the'' root of the chord, although Terhardt described
\emph{qualitative features} which can also alter the root of a chord (e.g.
partial $n$ is ``more important'' than partial $n+1$). Since these features are
not well defined, we will not described them in detail here (and  our
implementation ignores them). Table~\ref{tab:CmajTerhardt} shows how, again, $C$
is determined as the root of the chord $E_4-G_4-C_5$.

\begin{table}[h]
    \begin{center}
    \begin{tabular}{c c c c}
        Perfect unison      & $E$       & $G$      & $C$      \\
        Perfect fifth below & $A$       & $C$      & $F$      \\
        Major third below   & $C$       & $E\flat$ & $A\flat$ \\
        Major second above  & $F\sharp$ & $A$      & $D$      \\
        Major second below  & $D$       & $F$      & $B$      \\
        \hline
        2nd degree roots    & & $D,\; F, \;A$ & \\
        3rd degree root     & & $C$ & \\
    \end{tabular}
    \end{center}
    \caption{Analysis of a $Cmaj$ chord in its first inversion using Terhardt's model.}
\label{tab:CmajTerhardt}
\end{table}

\subsection{Richard Parncutt}

\cite{Parncutt1997} based his model on Terhardt's, but extended it in the
following ways:

He states that the partials are of different importance and assigns them
different \emph{weights}. To strengthen the bass as possible root, he adds an
additional arbitrary weight to the bass note of the chord. Finally he makes a
distinction between different tonalities (major or minor) and assigns each
interval for each tonality a different additional weight\footnote{Parncutt calls
this \emph{prevailing tonality}.}. The note with highest final weight is
determined as the root of the chord. In Table~\ref{tab:CmajParncutt} we can see
that in this model both $C$ and $E$ could be roots of the chord.

\begin{table}[h]
    \begin{center}
        \begin{tabular}{ccccccccccccc}
                                & $C$ & $C\sharp$ & $D$ & $D\sharp$ & $E$ & $F$
                                & $F\sharp$ & $G$ & $G\sharp$ & $A$ & $A\sharp$
                                & $B$ \\
            \hline
            Perfect unison      & 10 & & & & 10 & & & 10 & & & & \\
            Perfect fifth below & 5 & & & & & 5 & & & & 5 & & \\
            Major third below   & 3 & & & 3 & & & & & 3 & & & \\
            Major second below  & & & 2 & & & & 2 & & & 2 & & \\
            Major second above  & & & 1 & & & 1 & & & & & 1 & \\
            \emph{Tonality (Cmajor)}     & 33 &    & 10 & 1  & 17 & 15 & 2  & 24 & 1  & 11 &    & 5  \\
            \emph{Bass}         & & & & & 20 & & & & & & & \\
            \hline
            Overall Weights     & 51 &   & 13 & 4 & 47 & 21 & 4 & 34 & 4 & 18 & 1 & 5
        \end{tabular}
    \caption{Analysis of a $Cmaj$ chord in its first inversion using Parncutt's model.}
    \medskip
    \small
    The large weight that is given to the bass note ($E$) means this is tight race
    between $C$ and $E$. Parncutt states that values differing by less than
    5 should be regarded as the same, which actually makes it so that both
    notes can be seen as roots.
    \end{center}
\label{tab:CmajParncutt}
\end{table}

\section{A New Model}
\label{chap:SchmidModel}


\subsection{Assumptions and Representation}

For the development of the model we made several assumptions, which will be
described here on the representation of the example-chord
$C_4-E_4-G_4-B\sharp_5$:

\begin{description}
    \item[Enharmonic Equivalence] \hfill \\
        We ignore enharmonic equivalence as it is (in the equal temperament) just a
        difference in notation and does not
        contribute anything to the hearing. We simply choose one of the notations.\\
        $C_4-E_4-G_4-B\sharp_5 \quad \rightarrow \quad C_4-E_4-G_4-C_6$

    \item[Octaves] \hfill \\
        We hypothesize that the absolute height of a note is not relevant for
        the determination of chord roots and can therefor be ignored.\\
        $C_4-E_4-G_4-C_5 \quad \rightarrow \quad C-E-G-C$

    \item[Duplications] \hfill \\
        We decided to ignore duplications of chords, because we assume that for
        the perception of the chord it does not matter how many instruments
        play a certain note\footnote{We assume that all notes in the chord are
            played (about) equally loud. Playing a single note
            disproportionately loud might lead to the perception of a different
        root, but further research will have to be done on this topic.}.\\
        $C-E-G-C \quad \rightarrow \quad C-E-G$

    \item[Pitch Names] \hfill \\
        The pitches of a chord only have meaning in relation to each other. This
        means that for example $C-E-G$ and $D-F\sharp-A$ will have the same
        relative root: If we map them to their respective pitch classes and
        normalize them so that the first pitch class is $0$, they will both have
        root ``0''.
        (To get the name of the root, this mapping has to be inverted.)\\
        $C-E-G \quad \rightarrow \quad 0-4-7 \quad $ and also $ \quad D-F\sharp-A \quad \rightarrow \quad 0-4-7$
\end{description}

Note that all of these assumptions are also made by the other described
models\footnote{With the two exceptions that Parncutt gives the lowest note
additional weight and the pitch names are (possibly) important for stacking
thirds.}.



\subsection{Definition}

Schmid leans heavily on the method of stacking thirds and his model works as follows:


    For each note in a chord, stack the other notes on top of it so that the
    distance between this note and the highest note is minimal, but the distance
    between each two notes is at least 3 (\emph{Minor third}).
    The notes with the lowest minimal distance are the roots of the chords.

The distance 3 is used, because it represents the smallest interval (except the
unison) that human hearing still perceives as consonant. Smaller intervals are
usually rated as dissonant by test persons~\citep{Roederer2000}.

\subsubsection{Example}
We will show how this works on the following chord:
$G-B-D-F\sharp-G$\footnote{This chord is usually called $Gmaj7$.}

\begin{enumerate}
    \item Transform the chord into our developed representation:\\
        $G-B-D-F\sharp-G \quad \rightarrow \quad 0-4-7-11$
    \item For each note, stack the others above it with minimal distance 3 and
        measure the distances to the highest note:\\\\
        \begin{tabular}{|c|c|c|}
            \hline
            Assumed Root & Stacked with Minimal Distance (Normalized Pitch
            Classes) & Distance \\
            \hline
            0 ($G$)         &  $0\;(G)-4\;(B)-7\;(D)-11\;(F\sharp)$  &  11 \\
            4 ($B$)         &  $0\;(B)-3\;(D)-8\;(G)-17\;(F\sharp)$  &  17 \\
            7 ($D$)         &  $0\;(D)-5\;(G)-9\;(B)-16\;(F\sharp)$  &  16 \\
            11 ($F\sharp$)  &  $0\;(F\sharp)-5\;(B)-8\;(D)-13\;(G)$  &  13 \\
            \hline
        \end{tabular}
    \item We determine $G$ as the root of the chord, because its distance is 11
        which is smaller than 13, 16 and 17.
\end{enumerate}

\subsubsection{Ambiguous Examples}

These next two examples show that this determination is not always unambiguous.

\begin{itemize}
    \item First consider the chord $C-G\sharp-A\sharp$: According to Schmid
        its root can either be the $G\sharp$ or the $A\sharp$, because they both
        have distance 14.\\\\
        \begin{tabular}{|c|c|c|}
            \hline
            Assumed Root & Stacked with Minimal Distance (Normalized Pitch
            Classes) & Distance \\
            \hline
            0 ($C$)         &  $0\;(C)-10\;(A\sharp)-20\;(G\sharp)$  &  20 \\
            8 ($G\sharp$)   &  $0\;(G\sharp)-4\;(C)-14\;(A\sharp)$  &  14 \\
            10 ($A\sharp$)  &  $0\;(A\sharp)-10\;(G\sharp)-14\;(C)$  &  14 \\
            \hline
        \end{tabular}

    \item Now consider the chord $ E-G-A\sharp-C\sharp $: For each note the minimal
        distance is the same (9), because the pitch classes are always $0-3-6-9$,
        regardless of the lowest note! This is just one of several possible
        \emph{cyclic chords}\footnote{A cyclic chord is a chord in which the
        contained pitch classes all have the same distance from their neighbors
        when stacked using the shortest possible distance.}, for which nothing can be determined.\\\\
        \begin{tabular}{|c|c|c|}
            \hline
            Assumed Root & Stacked with Minimal Distance (Normalized Pitch
            Classes) & Distance \\
            \hline
            0 ($E$)         &  $0\;(E)-3\;(G)-6\;(A\sharp)-9\;(C\sharp)$  &  9 \\
            3 ($G$)         &  $0\;(G)-3\;(A\sharp)-6\;(C\sharp)-9\;(E)$  &  9 \\
            6 ($A\sharp$)   &  $0\;(A\sharp)-3\;(C\sharp)-6\;(E)-9\;(G)$  &  9 \\
            9 ($C\sharp$)   &  $0\;(C\sharp)-3\;(E)-6\;(G)-9\;(A\sharp)$  &  9 \\
            \hline
        \end{tabular}
\end{itemize}

\subsection{Interval Order}

\begin{figure}[h]
    \lilypondinclude{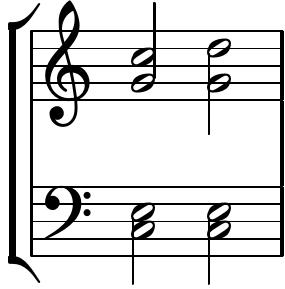}
    \caption{Interval order examples.}
    \begin{center}
        \medskip
        \small

    The first chord is a simple $Cmaj$ chord ($C-E-G-C$) with root $C$, while
    the root of the second chord ($C-E-G-D$) is ambiguous ($C$ or $D$). Since
    the interval  between $C$ (root of the first chord) and $C$ (first possible
    root of the second chord) is rated higher in the interval orders than the
    interval between $C$ and $D$ (Perfect unison (0) $>$ Major second (2)), the
    $C$ is chosen as the root for the second chord.

    \end{center}
\label{fig:IntervalOrdersExample}
\end{figure}

To be able to choose an unambiguous root from two or more ambiguous roots,
Schmid proposes to use an \emph{interval order}, in which the intervals are
ordered by ``importance''. The intervals to the previous and following chord
(i.e.~the distances between the pitch classes of the roots) have to be
determined and rated\footnote{It is not yet clear in what proportion the
previous and following chord influence the chord in the middle. For our
implementation we assumed that both intervals are equally important, but this
can easily be changed in the future.}.
Figure~\ref{fig:PossibleIntervalOrders} displays two possible interval orders
and Figure~\ref{fig:IntervalOrdersExample} shows them used in an example.

\begin{figure}[h]
    $$(a) \quad 0 \; > \; 7 \; > \; 5 \; > \; 4 \; > \; 8 \; > \; 3 \; > \; 9 \;
    > \; 2 \; > \; 10 \; > \; 1 \; > \; 11 \; > \; 6$$
    $$(b) \quad 0 \; > \; 7 \; = \; 5 \; > \; 4 \; = \; 8 \; > \; 3 \; = \; 9 \;
    > \; 2 \; = \; 10 \; > \; 1 \; = \; 11 \; > \; 6$$
    \caption{Two possible interval orders for the determination of unambiguous
    roots.}
    \small
    \begin{center}
    The numbers represent the number of semitones between the two compared pitch
    classes. Both orders rate the Perfect unison (or Perfect octave) the
    highest. The first order classifies the Perfect fifth above higher than the
    Perfect fifth below, while the second order values them with equal
    importance. Then this continues with the remaining intervals.
    \end{center}
\label{fig:PossibleIntervalOrders}
\end{figure}

\section{Chordification}
\label{chap:Chordification}

In order to determine the roots of chords in a pieces of music, we first need to
``combine'' the chords (from different voices).  Our definition of chords makes
this easy: Whenever one or more notes start or stop sounding, a new chord is
created (regardless of the voice the note appears in).

\begin{figure}[h!]
\begin{center}
    \lilypondinclude{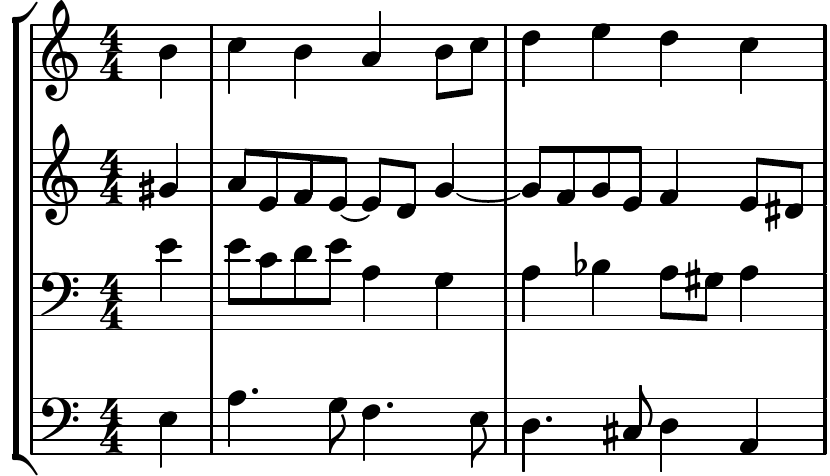}
    $ \wideDownarrow $
    \vspace{-0.7cm}
    \lilypondinclude{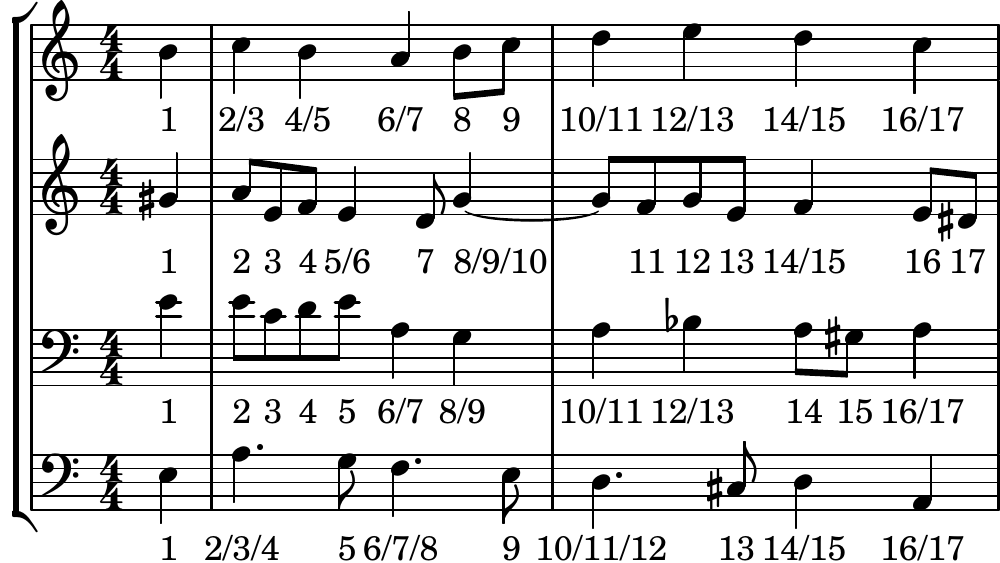}
    $ \wideDownarrow $
    \vspace{-0.7cm}
    \lilypondinclude{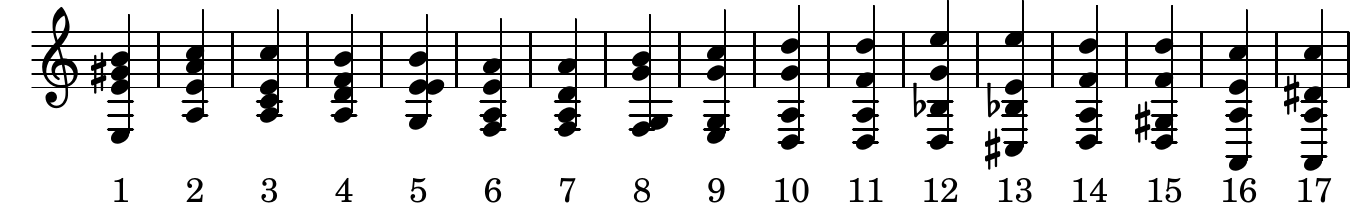}
    \caption{Chordification of a note sequence in two steps.}
    \medskip
    \small
\label{fig:ChordificationExample}
\end{center}
\end{figure}

Figure~\ref{fig:ChordificationExample} shows an example of this chordification
in two steps: First all the notes get assigned list of numbers indicating in
which chords they will sound in and then for each number a new chord is created.

\section{Example Analysis}

\begin{figure}[h]
    \lilypondinclude{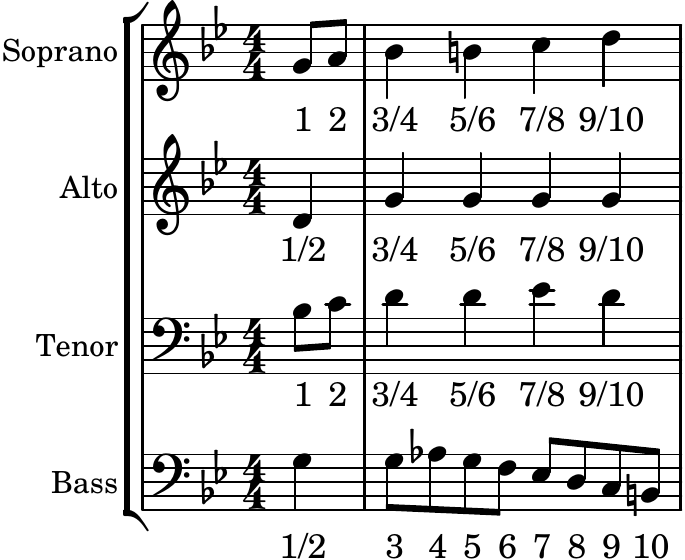}
    \medskip
    \begin{center}
    \begin{tabular}{r|c|c|c|c|c|c|c|c|c|c|}
        \cline{2-11}
                        & 1 & 2 & 3 & 4 & 5 & 6 & 7 & 8 & 9 & 10 \\
        \hline
        \multicolumn{1}{|r|}{Stacking Thirds}
        & $G$ & $A$ & $G$ & $G$ & $G$ & $G$ & $C$ & $C$ & $C$ & $G$ \\
        \multicolumn{1}{|r|}{Terhardt}
        & $C$ & $C$, $D$, $F$, $G$ & $C$ & $C$, $B\flat$ & $G$ & $G$ & $F$ & $C$, $F$ & $C$ & $G$ \\
        \multicolumn{1}{|r|}{Parncutt}
        & $G$ & $G$ & $G$ & $G$ & $G$ & $G$ & $E\flat$ & $D$ & $C$, $G$ & $G$ \\
        \multicolumn{1}{|r|}{Schmid}
        & $G$ & $A$ & $G$ & $G$ & $G$ & $G$ & $C$ & $D$ & $D$ & $G$ \\
        \hline
        \multicolumn{1}{|r|}{Correct Roots}
        & $G$ & $G$ & $G$ & $G$ & $G$ & $G$ & $C$ & $C$ & $G$ & $G$ \\
        \hline
    \end{tabular}
    \caption{An analysis of a short chord sequence by different models.}
    \medskip
    \small
    The ensemble staff above shows the beginning of BWV 14.5 (in \emph{G minor}) with numbered
    notes (so the reader can see the original score, but also easily construct the
    chords). The table below depicts the root(s) each model determined per chord.
    The correct roots were annotated by hand (This represents my own subjective
    intuition of what should be the ``correct root'' when taking nonharmonic
    tones into account.)
    We can see that all models predict some of the roots differently. Stacking
    thirds actually seems to come the closest to my predicted ``correct roots''.
    \end{center}
\label{fig:ExistingModelsExample}
\end{figure}

Now that we know how to combine chords and determine roots, we can look look at
a complete example of an analysis of (the beginning of) a real-world musical
piece in Figure~\ref{fig:ExistingModelsExample}.

\chapter{Musical Context}
\label{chap:MusicalContext}

A potential problem of Schmid's basic model is that in the model no distinction
is made between nonharmonic and harmonic tones for the determination of
roots\footnote{Actually no model that we know of makes such a distinction.}.  We will now introduce a
new model called \emph{Context}, which extends the Schmid model. It will try to
predict nonharmonic tones through the musical context and then ignore them for
the determination of roots. First, chord groups are formed out of chord
sequences and then it is shown how chords influence each other within their
groups. Note that we are only considering the Bach corpus (tonal music) in this
thesis.






\section{Chord Groups}

\begin{figure}[t]
\begin{center}
    \lilypondinclude{res/bwv312_chordified_beginning}
    $ \wideDownarrow $
    \vspace{-0.7cm}
    \lilypondinclude{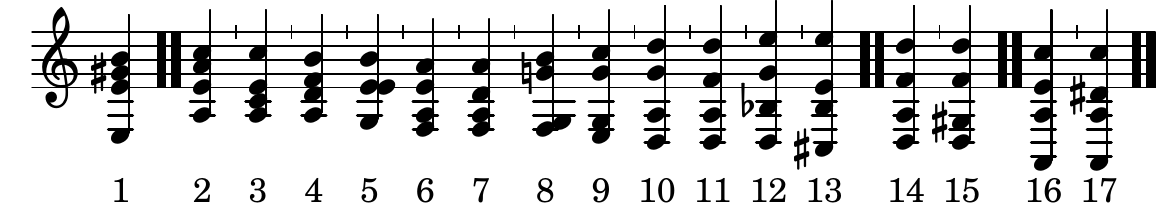}
    \caption{Creation of four different chord groups of different sizes.}
    \medskip
    \small
    \label{fig:ChordGroupsExample}
\end{center}
\end{figure}

First we introduce a new concept we call \emph{chord groups}, which are simply
sequences of chords:

In the beginning we put each chord in its own group and then, whenever two
consecutive chords share at least one note (i.e.~the note sounds in both
chords), we merge the two groups in which they are contained. Note that this
shared note actually needs to be the \emph{same} note and not two different
notes with equal pitch.  Figure~\ref{fig:ChordGroupsExample} shows an
example of this being done during the chordification we saw in
Figure~\ref{fig:ChordificationExample}. The black double bars delimit the
groups.

Within these groups it is now possible that neighboring chords influence each other
during the determination of roots: Chords may contain nonharmonic tones, which
gain meaning in combination with the previous or following chords. It is obvious that
with this definition, chords in chord groups of size 1 do not change.
From here on forth we will mean ``chord groups of size 2 or bigger'' when we
talk about chord groups.

\section{Features}
\label{chap:Features}

As a start we will look at chord groups of size 2, which we will call
\emph{chord pairs}. In order to be able to begin analysing them, we first need to
extract some features, which we will define in this section.

\subsubsection{Determined Root}

To each chord $c$ we assign a set $R(c)$ in which all its possible roots (or
rather the pitch classes of the roots) are contained. These roots are determined
by the Schmid model.


\subsubsection{Unique Root}
A chord can have a unique root or it can have multiple roots.
If a chord $c$ has a unique root we say $U(c)$ is true:
$$U(c) \quad \iff \quad size(R(c)) = 1$$

\subsubsection{Stack of Thirds}

As described earlier, the new model gives a lot of weight to \emph{minimal}
distances, but completely ignores \emph{maximal} distances.
Figure~\ref{fig:NicelyStackableFeatureExample} shows how these might actually be
useful:

\begin{figure}[h!]
\begin{center}
    \subfigure[$Cmaj$ with suspension]{
        \includegraphics{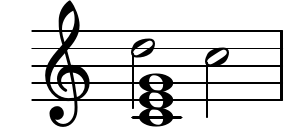}
    }
    \quad
    \subfigure[The first chord stacked on top of $D$]{
        \includegraphics{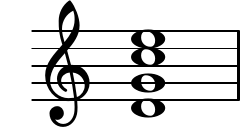}
    }
    \caption{An example where maximal distances are useful.}
\label{fig:NicelyStackableFeatureExample}
    \medskip
    \small
    The first chord in (a) is $C-E-G-D$ which contains the nonchord tone $D$,
    the second chord in (a) is a normal $Cmaj$ chord ($C-E-G-C$). The root of
    the second chord is $C$. For the root of the first chord both $C$ and $D$
    are predicted as possible roots, because they both a minimal distance of 14,
    even though the $C$ might make more sense if we disregard at $D$ as
    nonharmonic tone.
\end{center}
\end{figure}

To combat this we introduce a new feature $N(c)$ for chord $c$, which will
``limit'' the maximal distance between the root (lowest note) and highest note.
It is defined as
$$N(c) \quad \iff \quad d(c) \;\; \leq \;\; 4 \times (P(c)-1)$$
where $d(c)$ is the mentioned distance and $P(c)$ the number of different pitch
classes in the chord. This basically means that we restrict the distance between
each two notes to 4 (\emph{Major third}), which again leans heavily on the idea
of stacking thirds. If the distance is greater, this feature gives a first hint
that we should look at this chord more closely as it might for example contain
nonharmonic tones.

\subsubsection{Containment and Difference}

A chord $c_A$ (or set of roots $R(c_A)$) is contained in $c_B$ if all of the pitch
classes of the notes contained in $c_A$ are also contained in the pitch classes
of the notes of $c_B$.  If $c_A$ is contained in $c_B$ we write $c_A
\subseteq c_B$.  We say that two chords are equal if they are both contained in each
other. (E.g. $C_4-C_3-C_5-E_4-G_3-G_2 \; \; = \; \; C_4-E_4-G_4$)

Similarly, the difference of the chords ($c_A-c_B$) is defined as the set of all
notes the pitch classes of which are contained in $c_A$ but not in $c_B$.

\subsubsection{Partial Sub-Chord}
\begin{figure}[h]
\begin{center}
    \subfigure[First chord is fully contained]{
        \includegraphics{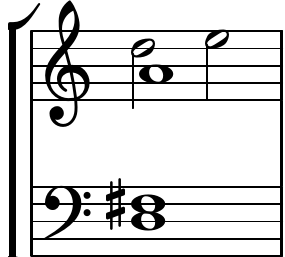}
\label{fig:PartialSubchordExample1}
    }
    \quad
    \subfigure[Second chord contains nonharmonic tone]{
        \includegraphics{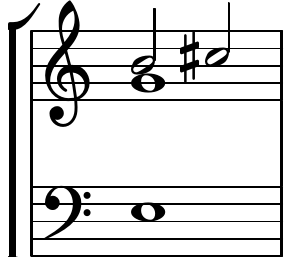}
\label{fig:PartialSubchordExample2}
    }
    \quad
    \subfigure[First chord contains nonharmonic tone]{
        \includegraphics{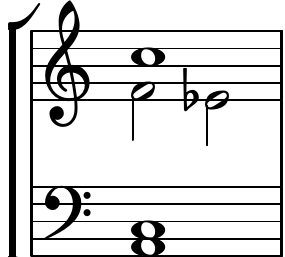}
    }
    \quad
    \subfigure[First chord might contain two nonharmonic tones]{
        \includegraphics{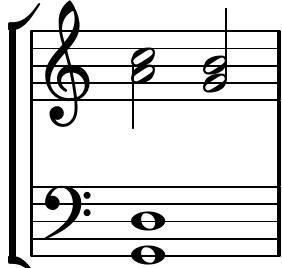}
    }
    \caption{Examples of partial sub-chords.}
\label{fig:PartialSubChordMotivation}
    \medskip
    \small
    \begin{description}

        \item[(a)] The first chord ($D-F\sharp-A-D$, predicted root $D$) is
            fully contained in the second chord ($D-F\sharp-A-E$, predicted
            roots $D$ or $E$) and is therefore a partial sub-chord.  If we look
            at the $E$ as a nonharmonic tone, it would probably not make sense
            to set it as the root of the chord. The partial sub-chord is
            supposed to help with this decision as it tells us that the root of
            the first chord might be more important and that we will set the
            root of the second chord to $D$.

        \item[(b)] The second condition comes into play: The root of the
            first chord ($E$) is contained in the second chord, but the root of
            the second chord ($C\sharp$) is not contained in the first chord. We
            will assume $C\sharp$ is a nonchord tone and set the root for the
            second chord to $E$.

        \item[(c)] Here the reverse of (b) happens: We will assume $F$ (the root
            of the first chord) is a nonharmonic tone and set the root to $A$
            (the root of the second chord) instead.

        \item[(d)] The first chord is $G-D-A-C$ with determined root $A$ and the
            second chord is $G-D-G-B$ with determined root $G$. We will take the
            (potentially controversial) step of assuming both $A$ and $C$ as
            nonchord tones and setting the root of the first chord to $G$. For
            pieces outside of the Bach corpus this might not work that well and
            additional restrictions will have to be developed. Note that the
            $H$-feature prevents a chord from having more nonharmonic tones than
            harmonic tones.

    \end{description}
\end{center}
\end{figure}

We say chord $c_A$ is a partial sub-chord of chord $c_B$ if one of the following
conditions holds:
\begin{itemize}
    \item $c_A$ is contained in $c_B$: $$c_A \subseteq c_B$$
    \item None of the roots\footnote{Most of the time these
        ``roots'' are actually just a single root.} of $c_B$ are contained in
        $c_A$ and the unique root\footnote{Semantically it would perhaps
        make sense to not restrict this to unique roots. The reason we decided
        against it is because then we would have to add the uniqueness-condition
        whenever we use this feature.}
        of $c_A$ is contained in $c_B$:\\
        $$c_A - R(c_B) = c_A \quad \land \quad R(c_A) \subseteq c_B \quad \land
        \quad U(c_A)$$
\end{itemize}
We write $c_A \unlhd c_B$ to express that $c_A$ is a partial sub-chord of $c_B$.
Figure~\ref{fig:PartialSubChordMotivation} depicts the motivation behind this
feature.

\begin{figure}[h!]
\label{fig:FutureWorkExample}
\begin{center}
    \subfigure[Two separate chords]{
        \includegraphics{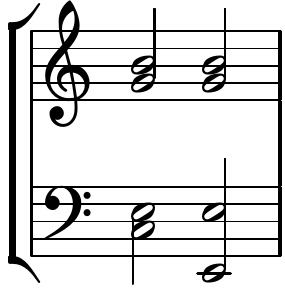}
    }
    \quad
    \subfigure[Bass note changes]{
        \includegraphics{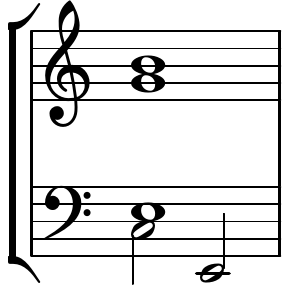}
    }
    \quad
    \subfigure[Highest note changes]{
        \includegraphics{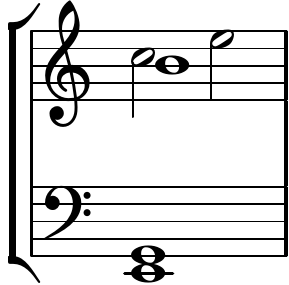}
    }
    \caption{Differences in changing tones in a chord pair.}
\label{fig:NumberOfChangingTonesProblem}
    \medskip
    \small
    In (a) we see the two separate chords $Cmaj7$ and $Emaj$ (which unlike in
    (b) and (c) do not form a chord pair). With our current representation we
    can not distinguish between the two progressions in (b) and (c). In (b) the
    bass note changes, which means the roots are probably $C$ and $E$, while in
    (c) the highest note changes which makes it more likely that the root is $E$
    for both chords~(Schmid, 2015, personal communication).
\end{center}
\end{figure}

\subsubsection{Number of Changing Notes}

We count the number of notes in chord $c$ that still sound from the previous
chord (which by definition is in the same chord group) and the number of notes
that only start sounding in the current chord. We say $H(c)$ holds, if there are
more ``new'' notes (that just started sounding) than ``old'' ones (which already
sounded in the previous chord). In short, $H(c)$ is true if ``more than half the
notes changed from the previous chord''. Note that most chords in our considered
Bach corpus consist of three or four different notes which reduces the decision
to ``Is at most a single note still sounding from the previous chord?''.

This again works for most of the Bach corpus, but it also reveals a potential
problem of our representation of chords, as can be seen in
Figure~\ref{fig:NumberOfChangingTonesProblem}, because we do not distinguish
between voices and ignore absolute pitches.

\section{Decision Tree for Chord Pairs}

Having determined useful features of our chord pairs, we can now start building
a decision tree. The input of the tree will be the arbitrary chord pair ($X$,
$Y$), complete with all the features that were just described, and the output
will be the two new (sets of) roots $R_X$ and $R_Y$.
The following sections will explain all the decisions that are made in the tree
and give examples of correctly determined roots for chord pairs made possible
by these decisions.

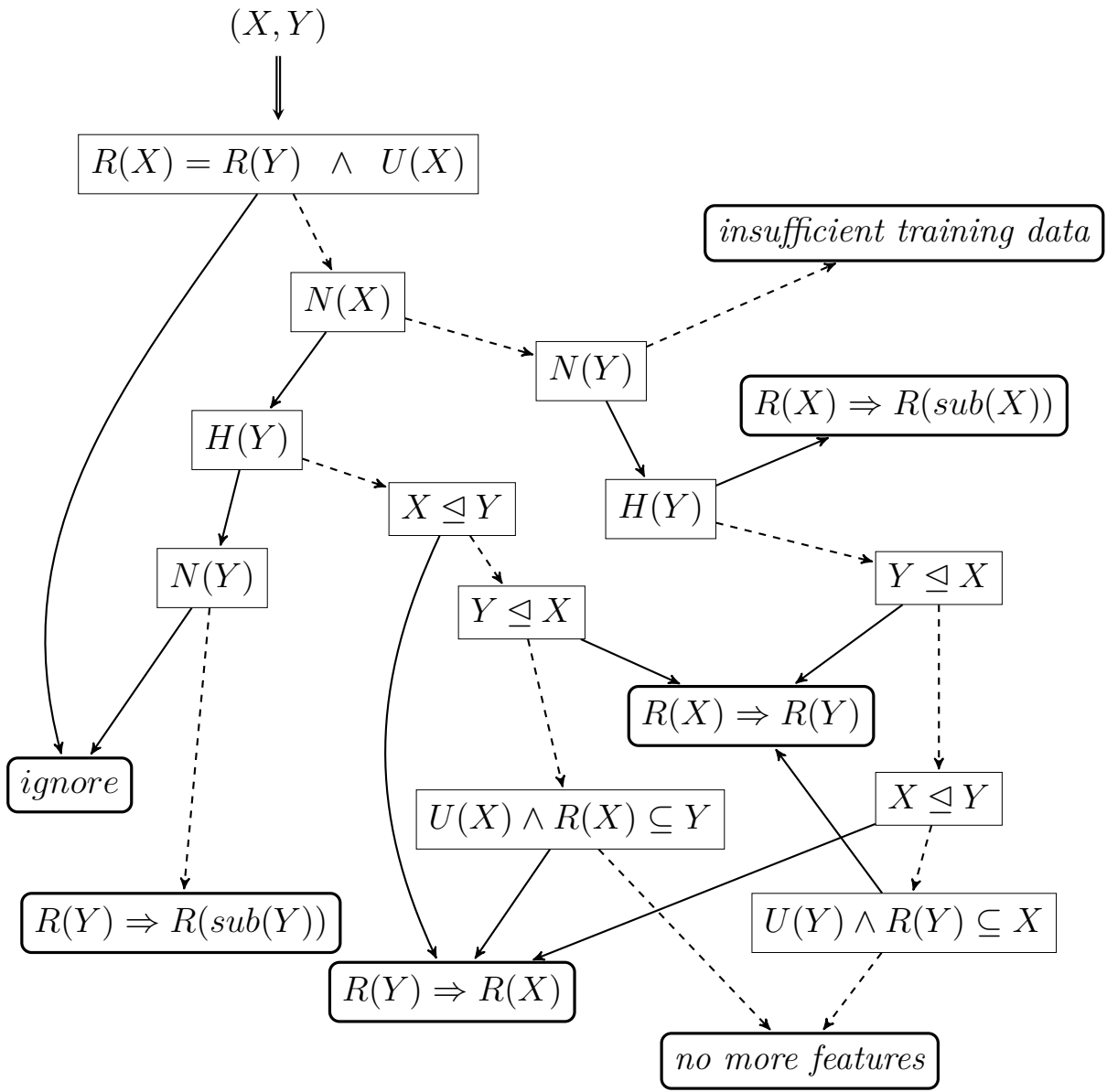
\begin{figure}[h!]
\begin{center}
\begin{tikzpicture}[->, >=stealth', shorten >=1pt, node distance=3cm, thick,
   label/.style={%
   postaction={ decorate,
   decoration={ markings, }}}]
    \tikzstyle{every node}=[draw, font=\Large, very thin, inner sep=5pt, minimum size=0mm]

    \node[draw=none] (input) at (0, 2) [] {$(X, Y)$};

    \node[rectangle] (0) at (0,0) {$R(X) = R(Y) \; \; \land \; \; U(X)$};

    \node (2) at (1, -2) [] {$N(X)$};
    \node (21) at (-0.45, -4) [] {$H(Y)$};
    \node (211) at (-0.95, -6) [] {$N(Y)$};
    \node (212) at (2.5, -5) [] {$X \unlhd Y$};
    \node (2122) at (3.5, -6.5) [] {$Y \unlhd X$};
    \node (22122) at (4.2, -9.5) [] {$U(X) \land R(X) \subseteq Y$};

    \node (22) at (4.5, -3) [] {$N(Y)$};
    \node (221) at (5.5, -5) [] {$H(Y)$};
    \node (2211) at (9.5, -6) [] {$Y \unlhd X$};
    \node (22111) at (9.5, -9.2) [] {$X \unlhd Y$};
    \node (22112) at (9, -11) [] {$U(Y) \land R(Y) \subseteq X$};

    \node (1) at (-3, -9) [rounded corners, very thick] {$ignore$};
    \node (2112) at (-1.4, -11) [rounded corners, very thick] {$R(Y) \Rightarrow R(sub(Y))$};
    \node (2121) at (2.5, -12) [rounded corners, very thick] {$R(Y) \Rightarrow R(X)$};
    \node (21221) at (6.8, -8) [rounded corners, very thick] {$R(X) \Rightarrow R(Y)$};
    \node (2212) at (9, -3.5) [rounded corners, very thick] {$R(X) \Rightarrow R(sub(X))$};
    \node (222) at (9, -1) [rounded corners, very thick] {\emph{insufficient
    training data}};
    \node (ignore) at (7.5, -13) [rounded corners, very thick] {\emph{no more
    features}};

    \path[every edge/.style={font=\Large}];

    \path[draw, ->, >=stealth, double] (input) to[] (0, 0.6);

    \path[draw] (0) to[out=-125, in=105] (1);
    \path[draw, dashed] (0) to[] (2);

    \path[draw] (2) -- (21);
    \path[draw, dashed] (2) -- (22);

    \path[draw] (21) -- (211);
    \path[draw, dashed] (21) -- (212);

    \path[draw] (211) to[] (1);
    \path[draw, dashed] (211) -- (2112);

    \path[draw] (212) to[out=-115, in=115] (2121);
    \path[draw, dashed] (212) -- (2122);

    \path[draw] (2122) -- (21221);
    \path[draw, dashed] (2122) -- (22122);

    \path[draw] (22122) -- (2121);
    \path[draw, dashed] (22122) -- (ignore);

    \path[draw] (22) -- (221);
    \path[draw, dashed] (22) -- (222);

    \path[draw] (221) to[] (2212);
    \path[draw, dashed] (221) -- (2211);

    \path[draw] (2211) to[] (21221);
    \path[draw, dashed] (2211) to[] (22111);

    \path[draw] (22111) -- (2121);
    \path[draw, dashed] (22111) -- (22112);

    \path[draw] (22112) -- (21221);
    \path[draw, dashed] (22112) -- (ignore);
\end{tikzpicture}
\end{center}
    \caption{The decision tree for an arbitrary chord pair $(X, Y)$.}
    \small
    \begin{center}

    The solid lines represent ``condition holds'', the dotted lines represent
    ``condition does not hold''.\\

    Note that there are three different endpoints where we do not change
    anything: If we reach the ``ignore''-endpoint on the left we are confident
    that ignoring the context is okay. If we reach the ``insufficient training
    data''-node at the top right corner, we do not know enough to make a
    sensible decision so we ignore it due to the lack of further meaningful
    features and the fact that very few samples from our corpus reach this
    point. At the bottom ``no more features''-node we know that most
    features do not hold, which again forces us to ignore the pair. The
    following sections will explain the decisions that lead to these node more
    in-depth.

    \end{center}

\label{fig:DecisionTree}
\end{figure}

The whole tree in the form of a graph is displayed in
Figure~\ref{fig:DecisionTree}.

\subsection{Same Unique Root}

We start by filtering out all the pairs that already have the same unique root
($R(X) = R(Y) \; \; \land \; \; U(X)$).
This is done because we assume that the Schmid model is (in most cases) able to determine
the correct roots for isolated chords and that it is very likely for chords in
their chord pair to have the same root.
The two most prominent cases in which this rule is helpful are
\begin{itemize}
    \item The chords are equal
    \item One chord is contained in the other one, which contains an additional
    (often harmonic) tone (e.g.~the seventh of a triad)
\end{itemize}

\begin{figure}[h]
\begin{center}
    \subfigure[Equal chords]{
        \includegraphics{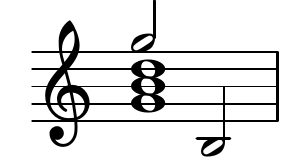}
    }
    \subfigure[Second chord contains $7th$]{
        \includegraphics{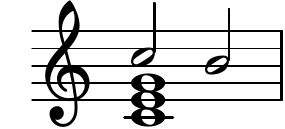}
    }
    \caption{Chord pairs where both chords have the same unique root.}
\label{fig:SameUniqueRootExamples}
\end{center}
\end{figure}

\noindent These pairs are simply ignored and their determined roots are not
changed. Figure~\ref{fig:SameUniqueRootExamples} shows two chord pairs to which
this condition applies.


\subsection{$X$ is Stack of Thirds}
\label{chap:Branch21}

As the section title suggests, we now decide whether or not $X$ is a stack of
thirds. If it is not, it might contain at least one nonharmonic tone. If $X$ is
a stack of thirds, there is a good chance that we already determined its root
correctly, but in both cases we need some more information.
\begin{itemize}
    \item If $N(X)$, we calculate if more than half the notes change from $X$ to
    $Y$
        \begin{itemize}
            \item $H(Y) \rightarrow$ continues immediately in Section~\ref{chap:Branch211}
            \item $\lnot H(Y) \rightarrow$ continues in Section~\ref{chap:Branch212}
        \end{itemize}
    \item If $\lnot N(X)$, we will make the next decision in Section~\ref{chap:Branch22}
\end{itemize}


\subsection{More Than Half the Notes Change}
\label{chap:Branch211}

$H(Y)$ essentially tells us that $X$ and $Y$ are only loosely connected and
most of their notes do not influence each others root.

\begin{figure}[h]
\begin{center}
    \subfigure[BWV 244.15]{
        \includegraphics{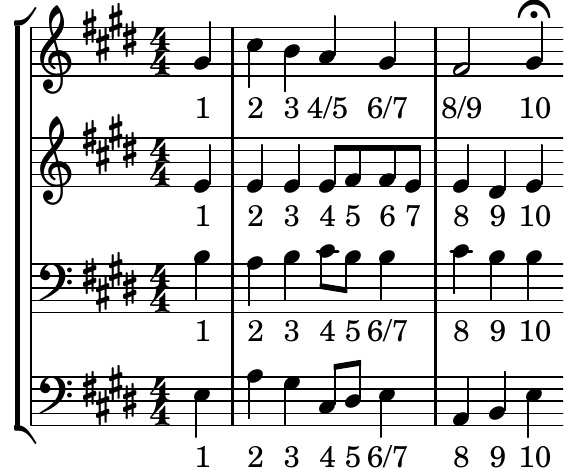}
        \label{fig:Branch211Example:subfig:1}
    }
    \subfigure[BWV 13.6]{
        \includegraphics{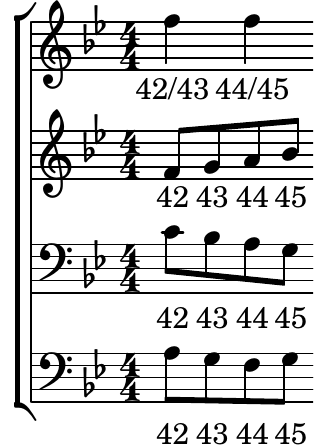}
        \label{fig:Branch211Example:subfig:2}
    }
    \caption{Chord pairs where more than half the note change.}
    \medskip
    \small

\begin{description}

    \item[(a)] This depicts the first case
        where $N(Y)$ holds: The chord with number 4 is $C\sharp-C\sharp-E-A$ and
        the next chord is $D\sharp-B-F\sharp-A$ (only $A$ keeps sounding). Both
        are stacks of thirds and the roots, $A$ and $B$ respectively, are
        determined without context. The chord pair 8--9 also falls in this
        category

    \item[(b)] We see that both chord
        pairs 42--43 and 44--45 have $\lnot N(Y)$. The $F$ is removed from both
        chord 43 and chord 45, which leaves us with the root $G$ in both cases.
        Chords 42 and 44 both have root $F$.

\end{description}

\label{fig:Branch211Example}
\end{center}
\end{figure}

In this branch we have only one last decision to make: Is $Y$ is also a stack of
thirds? If it is, we again leave both roots unchanged (although they could be
different now). If $\lnot N(Y)$ we have to do more work, because we know that $Y$
probably contains nonharmonic tones which still sound from $X$. The solution is
to remove from Y all those notes of X that still sound (less than half the
notes) and determine the roots of this subset of $Y$ without context.
Figure~\ref{fig:Branch211Example} shows examples of this branch.

\subsection{Not More Than Half the Notes Change}
\label{chap:Branch212}

If only less than half or half the notes change, we know that the two chords are more
closely related and it is probable that they have the same root. We now check if
one of the partial sub-chord conditions holds:
\begin{itemize}
    \item If $X \unlhd Y$, we set the root of $Y$ to that of $X$
    \item If $\lnot (X \unlhd Y)$, we check the reverse:
        \begin{itemize}
            \item If $Y \unlhd X$, we set the root of $X$ to that of $Y$
            \item If $\lnot (Y \unlhd X)$, but $U(X)$ and $R(X) \subseteq Y$,
                we again set the root of $Y$ to that of $X$
            \item Otherwise we leave both roots unchanged
        \end{itemize}
\end{itemize}

\begin{figure}[h!]
\begin{center}
    \subfigure[$X \unlhd Y$]{
        \includegraphics{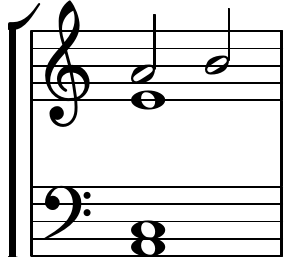}
    }
    \subfigure[$Y \unlhd X$]{
        \includegraphics{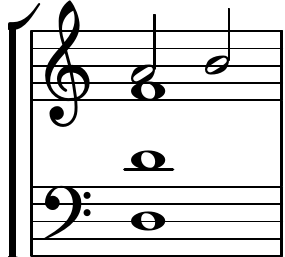}
    }
    \subfigure[$U(X) \land R(C) \subseteq Y$]{
        \includegraphics{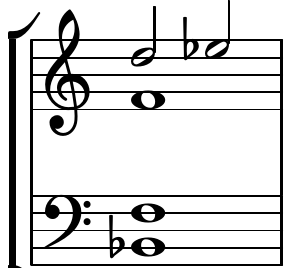}
    }
    \subfigure[Leave unchanged]{
        \includegraphics{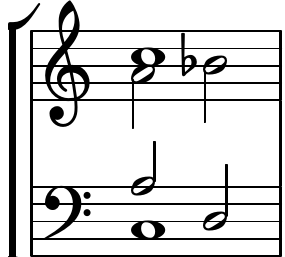}
    }
    \caption{Chord pairs where $N(X)$ and not more than half the notes change.}
    \medskip
    \small

\begin{description}

    \item[(a)] $X$ is contained in $Y$ and therefore also a partial sub-chord of
        $Y$. We set the root of $Y$ to $A$. A similar example was already shown
        in Figure~\ref{fig:PartialSubchordExample1}.

    \item[(b)] The predicted root of $Y$ ($D-D-F-B$) is $B$, which is not
        contained in chord $X$ ($D-D-F-A$), but the root of $X$ ($D$) is
        contained in $Y$, so the root of $Y$ is set to $D$. Again, we have
        already seen a similar example in
        Figure~\ref{fig:PartialSubchordExample2}.

    \item[(c)] Both chords are not partial sub-chords of each other. The
        predicted root for $X$ is $B\flat$, which is unique and contained in
        $Y$, while $Y$ has the predicted root $F$, which is not contained in
        $X$. We determine the root of $Y$ as $B\flat$.

    \item[(d)] The predicted root for $X$ ($C-A-A-C$) is $A$, while the
        predicted root for $Y$ ($C-D-B\flat-C$) is ambiguous: either $C$ or
        $B\flat$. Both chords are not partial sub-chords of each other and the
        root of $X$ is not contained in $Y$, so we leave the roots unchanged.
        This means we were unable to determine an unambiguous root for chord
        $Y$.

\end{description}

\label{fig:Branch212Example}
\end{center}
\end{figure}

Figure~\ref{fig:Branch212Example} shows example of these branches.

\subsection{$Y$ is Stack of Thirds}
\label{chap:Branch22}

The branch in which $N(Y)$ holds will almost mirror the decisions in
Section~\ref{chap:Branch21}, with the exception that we already know $\lnot
N(X)$. Again we first check if more than half the notes change and then if one of
the partial sub-chord conditions holds.

\begin{figure}[h!]
\begin{center}
    \subfigure[Root of the first chord changes]{
        \includegraphics{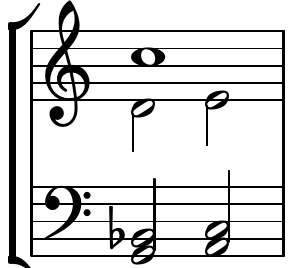}
    }
    \quad
    \subfigure[Root does not change]{
        \includegraphics{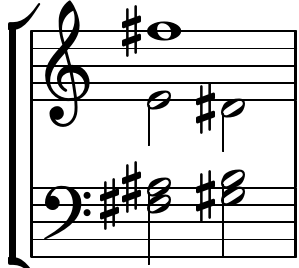}
    }
    \caption{Chord pairs where more than half the notes change.}
    \medskip
    \small

\begin{description}

    \item[(a)] The first chord is $G-B\flat-D-C$ with predicted roots $C$ or
        $B\flat$ and is not a stack of thirds. The second chord is $A-C-E-C$
        with root $A$ and is a stack of thirds. As we can see, only a single
        note still sounds in the second chord and we therefore remove that note
        from the first chord. With $sub(X) = G-B\flat-D$ we get the unambiguous
        root $G$ for $X$.

    \item[(b)] Sometimes the root of a chord does not change when we take out a
        note. We remove the highest $F\sharp$ from the first chord, but there is
        still the $F\sharp$ from the bass and its root remains $F\sharp$. The
        root of the second chord also remains $G\sharp$.

\end{description}
\label{fig:Branch22Example}
\end{center}
\end{figure}

\begin{itemize}
    \item If $H(Y)$, we assume the notes that still sound in $Y$ only belong to
    $Y$ and remove them from $X$ to determine the roots of this subset of $X$
    without context. Figure~\ref{fig:Branch22Example} shows examples of chord
    pairs where this holds.
    \item If $\lnot H(Y)$, we continue in Section~\ref{chap:Branch2211}.
\end{itemize}

\subsection{Partial Sub-Chords Mirrored}
\label{chap:Branch2211}

\begin{itemize}
    \item If $Y \unlhd X$, we set the root of $X$ to that of $Y$
    \item If $\lnot (Y \unlhd X)$, we again check the reverse:
        \begin{itemize}
            \item If $X \unlhd Y$, we set the root of $Y$ to that of $X$
            \item If $\lnot (X \unlhd Y)$, but $U(Y)$ and $R(Y) \subseteq X$,
                we again set the root of $X$ to that of $Y$
            \item Otherwise we leave both roots unchanged
        \end{itemize}
\end{itemize}

\subsection{No Stack of Thirds}
\label{chap:Branch222}

\begin{figure}[h!]
\begin{center}
    \subfigure[Roots are predicted unambigously]{
        \includegraphics{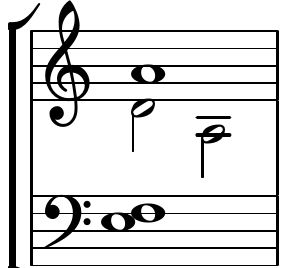}
    }
    \quad
    \subfigure[Ambiguous Roots]{
        \includegraphics{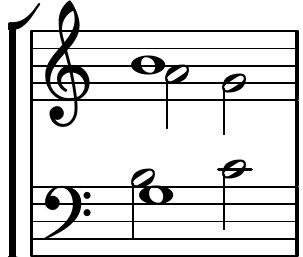}
    }
    \caption{Chord pairs without a stack of thirds.}
    \medskip
    \small

\begin{description}

    \item[(a)] For the first chord the $E$ is predicted as root, and for the
        second chord the $F$. In both cases we have a rather dissonant interval
        in the bass voice.

    \item[(b)] The root determination of the first chord is ambiguous: either
        $G$ or $A$. For the second chord we determine $C$ as the root (which
        means we assume that the third is missing from the chord, but fifth and
        seventh are present).

\end{description}

\label{fig:Branch222Example}
\end{center}
\end{figure}

This is where some special chords appear.
There are simply not enough examples of that in the Bach corpus we used
to come up with meaningful rules beyond just accepting the roots that were
determined without context. Two examples are displayed in
Figure~\ref{fig:Branch222Example}.

\subsection{Possible Outcomes}
\label{chap:PossibleOutcomes}

To summarize, we have five possible outcomes for the arbitrary chord pair ($X$, $Y$):
\begin{itemize}
    \item Do not change the previously determined roots for either chord
    \item Set the root of $X$ to that of $Y$
    \item Set the root of $Y$ to that of $X$
    \item Change the root of $X$ by removing some notes from $X$ and determining
    the root again without context
    \item Change the root of $Y$ by removing some notes from $Y$ and determining
    the root again without context
\end{itemize}

\section{Larger Chord Groups}

For larger groups (i.e.~groups with size larger than 2) we only use a single
rule as a first approach\footnote{This will likely have to be revisited and
made more complex after further research.}.

\subsection{Reduction to Pairs}

\begin{figure}[h!]
\begin{center}
    \subfigure[BWV 429]{
        \includegraphics{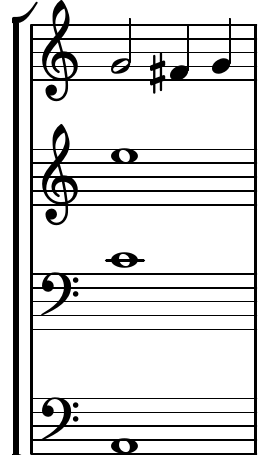}
    }
    \quad
    \subfigure[Group of length 3 (2 pairs)]{
        \includegraphics{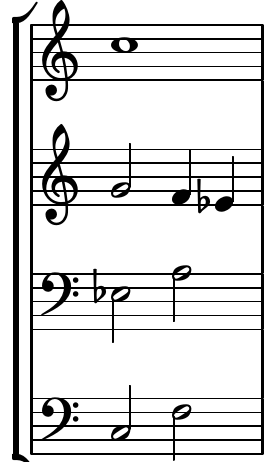}
    }
    \quad
    \subfigure[BWV 6.6]{
        \includegraphics{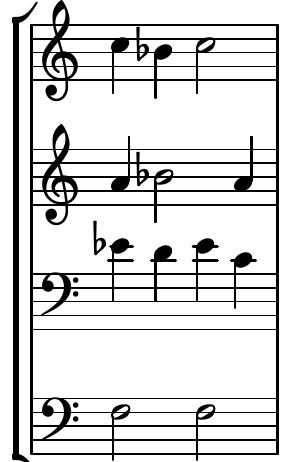}
    }
    \caption{A reduction of a large chord group to pairs.}
    \medskip
    \small

\begin{description}

    \item[(a)] The first and third chord consist of exactly the same notes,
        while the second chord contains the nonharmonic tone $F\sharp$ (which is
        also determined as the root). This is spotted and the root of the second
        chord is set to $A$, the root of the other two chords
        (c.f.~Section~\ref{chap:Branch212}).

    \item[(b)] The second group is reduced to the two pairs (the asterisks mark
        notes that did not change):
        \begin{itemize}
            \item $C-E\flat-G-C$ (root $C$) and $F-A-F-C^*$ (root $F$)\\
                We leave both roots unchanged
                (c.f.~Section~\ref{chap:Branch211}).
            \item $F-A-F-C^*$ (root $F$) and $F^*-A^*-E\flat-C^*$ (root $F$ or
                $E\flat$)\\
                We determine $F$ as the root of the second chord in this pair
                (c.f.~Section~\ref{chap:Branch212}).
        \end{itemize}
        Now we check if the predicted roots for the same chord in different
        pairs are equal. In this case we just have to check the second chord
        which has root $F$ in both pairs and we determine the roots $C$, $F$ and
        $F$ for this group.

    \item[(c)] Here we have 3 pairs:
        \begin{itemize}
            \item $F-E\flat-A-C$ (root $F$) and $F^*-D-B\flat-B\flat$ (root $B\flat$)\\
                We do not change either root (c.f.~Section~\ref{chap:Branch211}).
            \item $F^*-D-B\flat-B\flat$ (root $B\flat$) and $F-E\flat-B\flat^*-C$ (root $C$)\\
                We change the root of the second chord to $F$
                (c.f.~Section~\ref{chap:Branch211}).
            \item $F-E\flat-B\flat^*-C$ (root $C$) and $F^*-C-A-C^*$ (root $F$)\\
                We change the root of the first chord to $F$
                (c.f.~Section~\ref{chap:Branch2211}).
        \end{itemize}
        Again we check if the changed roots are the same, which is the case for
        the single changed root of chord 3 as it is set to $F$ in both pairs. We
        predict the roots $F$, $B\flat$, $F$ and $F$ for this chord group.

\end{description}

\label{fig:ReductionToPairsExample}
\end{center}
\end{figure}

We split the group into pairs. All the chords in the group except the first and
the last one will be part of exactly two groups (once as the first chord and
once as the second chord). Then we run all the pairs through the decision tree
and check if for each chord the results are the same in its two groups. Whenever
the determined roots for a chord are the same, we set the root of this chord to
this determined root. If the roots are different, we keep the root that was
computed without any context.
Figure~\ref{fig:ReductionToPairsExample} shows how this is done on chord groups
of length 3 and 4.

\chapter{Generation of a Decision Tree}
\label{chap:AutomaticDecisionTree}

The biggest challenge in creating a classifier for chord pairs was to find
meaningful features. The decision tree that was introduced in the previous
chapter was then created by hand. We also generated such a tree automatically,
which will be described in this chapter.

\section{Training}

For the creation of the decision tree we used the Python library
scikit-learn~\citep{scikit-learn2011}.  Gini impurity was used as the error
measure.  To prevent overfitting, we set the parameter ``\verb|min_samples_leaf=10|'' so that
it would stop splitting if less than 10 sample would be in the split branches.
The training data consisted of 613 chord pairs\footnote{Note that only
    \emph{pairs} are included here and not chords that are in their
own or larger groups.} $(X, Y)$ of which we manually
annotated the following 8 features\footnote{We described these features in
Section~\ref{chap:Features}.} (with either ``true'' or ``false''):

\begin{itemize}
    \item $N(X)$
    \item $N(Y)$
    \item $X \unlhd Y$
    \item $Y \unlhd X$
    \item $R(X) = R(Y) \; \land \; U(X)$
    \item $H(Y)$
    \item $U(X) \land R(X) \subseteq Y$
    \item $U(Y) \land R(Y) \subseteq X$
\end{itemize}

\noindent In addition to that each pair was annotated with a class, which
denoted the suspected outcome of its analysis (which were already listed in
Section~\ref{chap:PossibleOutcomes}):
\begin{itemize}
    \item Do not change the previously determined roots for either chord
    \item Set the root of $X$ to that of $Y$
    \item Set the root of $Y$ to that of $X$
    \item Change the root of $X$ by removing some notes from $X$ and determining
    the root again without context
    \item Change the root of $Y$ by removing some notes from $Y$ and determining
    the root again without context
\end{itemize}

As only restriction we chose to stop splitting when a split would create a node
with less than 10 training samples to avoid over-fitting.

\section{Outcome}

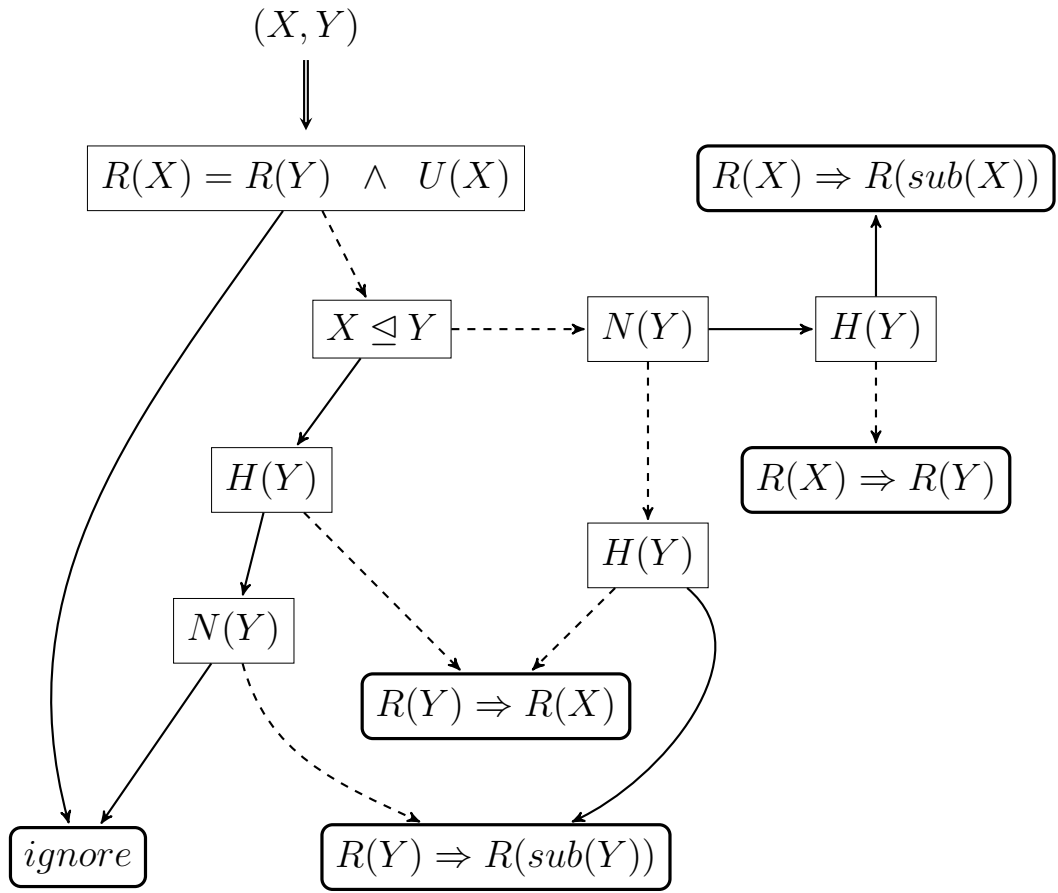
\begin{figure}[h!]
\begin{center}
\begin{tikzpicture}[->, >=stealth', shorten >=1pt, node distance=3cm, thick,
   label/.style={%
   postaction={ decorate,
   decoration={ markings, }}}]
    \tikzstyle{every node}=[draw, font=\Large, very thin, inner sep=5pt, minimum size=0mm]

    \node[draw=none] (input) at (0, 2) [] {$(X, Y)$};
    \path[draw, ->, >=stealth, double] (input) to[] (0, 0.6);

    \node[rectangle] (0) at (0,0) {$R(X) = R(Y) \; \; \land \; \; U(X)$};

    \node (2) at (1, -2) [] {$X \unlhd Y$};
    \node (21) at (-0.45, -4) [] {$H(Y)$};
    \node (211) at (-0.95, -6) [] {$N(Y)$};

    \node (22) at (4.5, -2) [] {$N(Y)$};
    \node (221) at (7.5, -2) [] {$H(Y)$};
    \node (222) at (4.5, -5) [] {$H(Y)$};
    \node (ignore) at (-3, -9) [rounded corners, very thick] {$ignore$};
    \node (YtoX) at (2.5, -7) [rounded corners, very thick] {$R(Y) \Rightarrow R(X)$};
    \node (XtoY) at (7.5, -4) [rounded corners, very thick] {$R(X) \Rightarrow R(Y)$};
    \node (subX) at (7.5, 0) [rounded corners, very thick] {$R(X) \Rightarrow R(sub(X))$};
    \node (subY) at (2.5, -9) [rounded corners, very thick] {$R(Y) \Rightarrow R(sub(Y))$};

    \path[draw] (0) to[out=-125, in=105] (ignore);
    \path[draw, dashed] (0) to[] (2);

    \path[draw] (2) -- (21);
    \path[draw, dashed] (2) -- (22);

    \path[draw] (21) -- (211);
    \path[draw, dashed] (21) -- (YtoX);

    \path[draw] (211) to[] (ignore);
    \path[draw, dashed] (211) to[out=-75, in=155] (subY);

    \path[draw] (22) -- (221);
    \path[draw, dashed] (22) -- (222);

    \path[draw] (221) to[] (subX);
    \path[draw, dashed] (221) -- (XtoY);

    \path[draw] (222) to[out=-40, in=25] (subY);
    \path[draw, dashed] (222) -- (YtoX);

\end{tikzpicture}
\end{center}
    \caption{Automatically created decision tree for an arbitrary chord pair $(X, Y)$.}
    \small
    \begin{center}

    The solid lines represent ``condition holds'', the dotted lines represent
    ``condition does not hold''.\\
    Note that the tree only uses 4 of the 8 possible features.

    \end{center}

\label{fig:GeneratedDecisionTree}
\end{figure}

The generated tree is displayed in Figure~\ref{fig:GeneratedDecisionTree}.

The tree yielded a 96.23\% accuracy on the test-data (consisting of 106 samples,
which were taken out of the original data before creating the tree\footnote{We
first extracted every 10th sample then again every 22nd sample from the original
data.}). When running on our corpus of 26 manually annotated musical pieces from
Johann Sebastian Bach\footnote{Note that the training samples are included in
there.}, the generated tree was able to determine 85.01\% of the roots
correctly, while the manually created tree achieves an accuracy of
95.34\%.\footnote{The original Schmid model scores 80.21\%.
Appendix~\ref{chap:AppendixBenchmarks} contains a detailed list of benchmarks
for all models.} The
tree we created by hand also makes use of more features and we think that it
makes a bit more sense from a music theoretical standpoint (E.g.~for the outcome
$R(Y) \Rightarrow R(X)$ it is actually important that $\lnot N(X)$, which is
checked in our manual tree, but not in the generated tree.). The automatic tree also
does not ignore data, when not enough features hold.
Our program includes the code for both trees, but uses the manually generated
one by default.

\chapter{Implementation}
\label{chap:Implementation}

In this chapter we give a short overview of the capabilities of our program and
the technologies we used. Then we give a brief introduction on how to use the
program via both command line and graphical user interface.

In general, readability and extensibility were favored over efficiency in our
implementation (e.g.~roots of some of the models could be calculated more
quickly, but the ideas of the models would not be obvious anymore and the speed
gain would probably be imperceptible).

%
%

\section{Capabilities}

Our program is able to
\begin{itemize}

    \item Read in musical pieces (either a single piece or all pieces in a
    certain directory) encoded in various file formats.

    \item Analyse these pieces using the following models:
    \begin{itemize}
        \item Stacking Thirds (from the music21 library; see
            Section~\ref{chap:UsedTechnologies})
        \item Terhardt
        \item Parncutt
        \item Schmid (with and without interval order)
        \item Context
    \end{itemize}

    \item Output the following files: 
    \begin{itemize}

        \item HTML file containing the analysis for each chord from the analysed
            musical piece

        \item MusicXML file containing the original piece with the number(s) of
            each chord annotated in the `lyrics'\footnote{We already saw an
            example of that in the middle part of
            Figure~\ref{fig:ChordificationExample}.}

        \item Textfiles for each model containing the determined roots for the
            chords (one per line)

    \end{itemize}
\end{itemize}

\section{Used Technologies}
\label{chap:UsedTechnologies}

The main programming language that was used is
Python~3\footnote{\cite{python3}}. Our code depends heavily on the
music21--library\footnote{\cite{music21}} for things such as parsing the input
files or writing out the numbered MusicXML files. The GUI was created with
PyQT5\footnote{\cite{pyqt5}}. The HTML files include some JavaScript which uses
the libraries jQuery\footnote{\cite{jquery}} and
VexFlow\footnote{\cite{vexflow}} (to display staves and notes). The CSS
framework Bootstrap\footnote{\cite{bootstrap}} was used to improve formatting
and make the tables look nice.

%
%
%
%
\section{Example Usage}
\label{chap:ExampleUsage}

\subsection{Command Line}

Our program was mainly developed for command line usage.
The simplest possible possible usecase is to analyse a single file.

\begin{lstlisting}[basicstyle=\ttfamily]
  $ python3 analyse.py music.mxl
  Analysing: music.xml
\end{lstlisting}

%

\noindent This will do an analysis of the file with all the models and write out
an HTML file of the results with name ``music.html''.
We can also do bulk analysis of all files contained in directory. In this next
example we also set the ``verbose'' flag to get more information and the
``musicxml'' flag to output numbered MusicXML files.

\begin{lstlisting}[basicstyle=\ttfamily]
  $ python3 analyse.py music/pieces/ -v --musicxml
  Analysing: music/pieces/music_1.mxl
  Wrote HTML file (music/pieces/music_1.html)
  Wrote MusicXML file (music/pieces/music_1.numbered.xml)
  Done!

  Analysing: music/pieces/music_2.mxl
  Wrote HTML file (music/pieces/music_2.html)
  Wrote MusicXML file (music/pieces/music_2.numbered.xml)
  Done!
  ...
\end{lstlisting}

\noindent The complete list of possible options can be found in
Appendix~\ref{chap:AppendixProgramOptions}.

\subsection{GUI}

\begin{figure}[h]
    \begin{center}
    \scalebox{0.35}[0.35]{
    \includegraphics{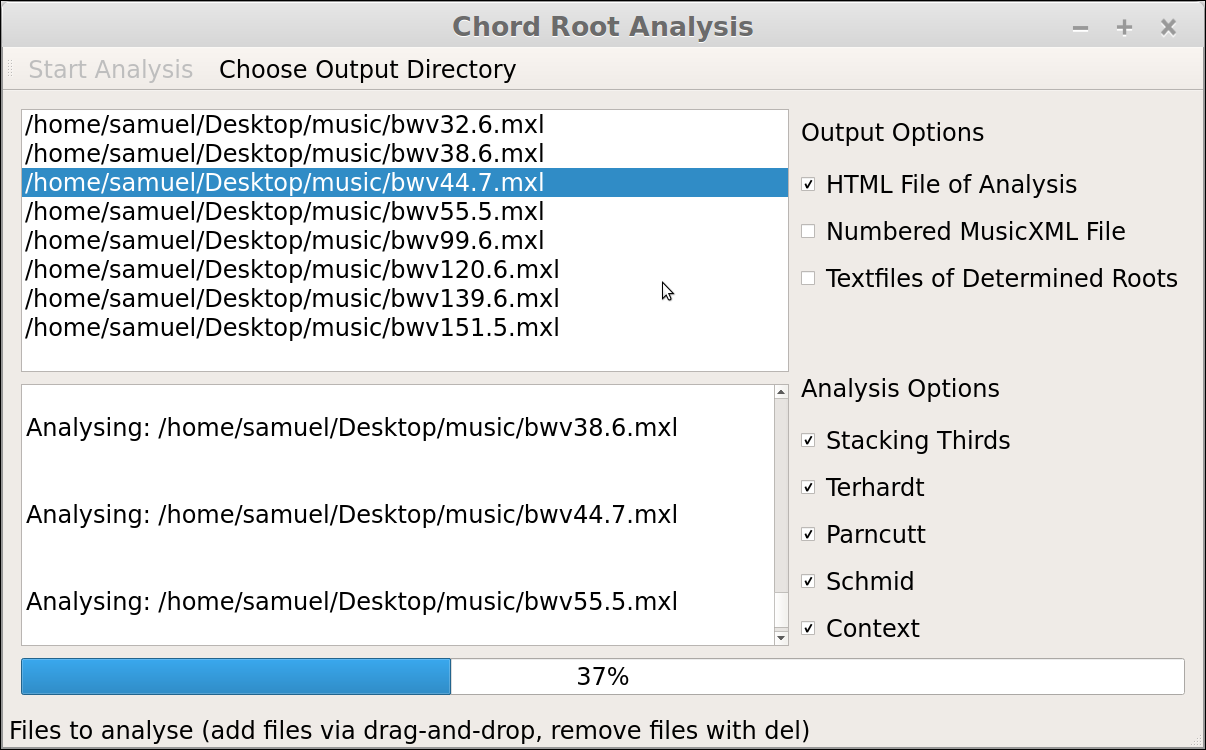}
}
    \end{center}
\label{fig:GUIExample}
    \caption{A screenshot of the program during analysis.}
\end{figure}

\begin{figure}[h]
    \begin{center}
    \scalebox{0.22}[0.22]{
        \includegraphics{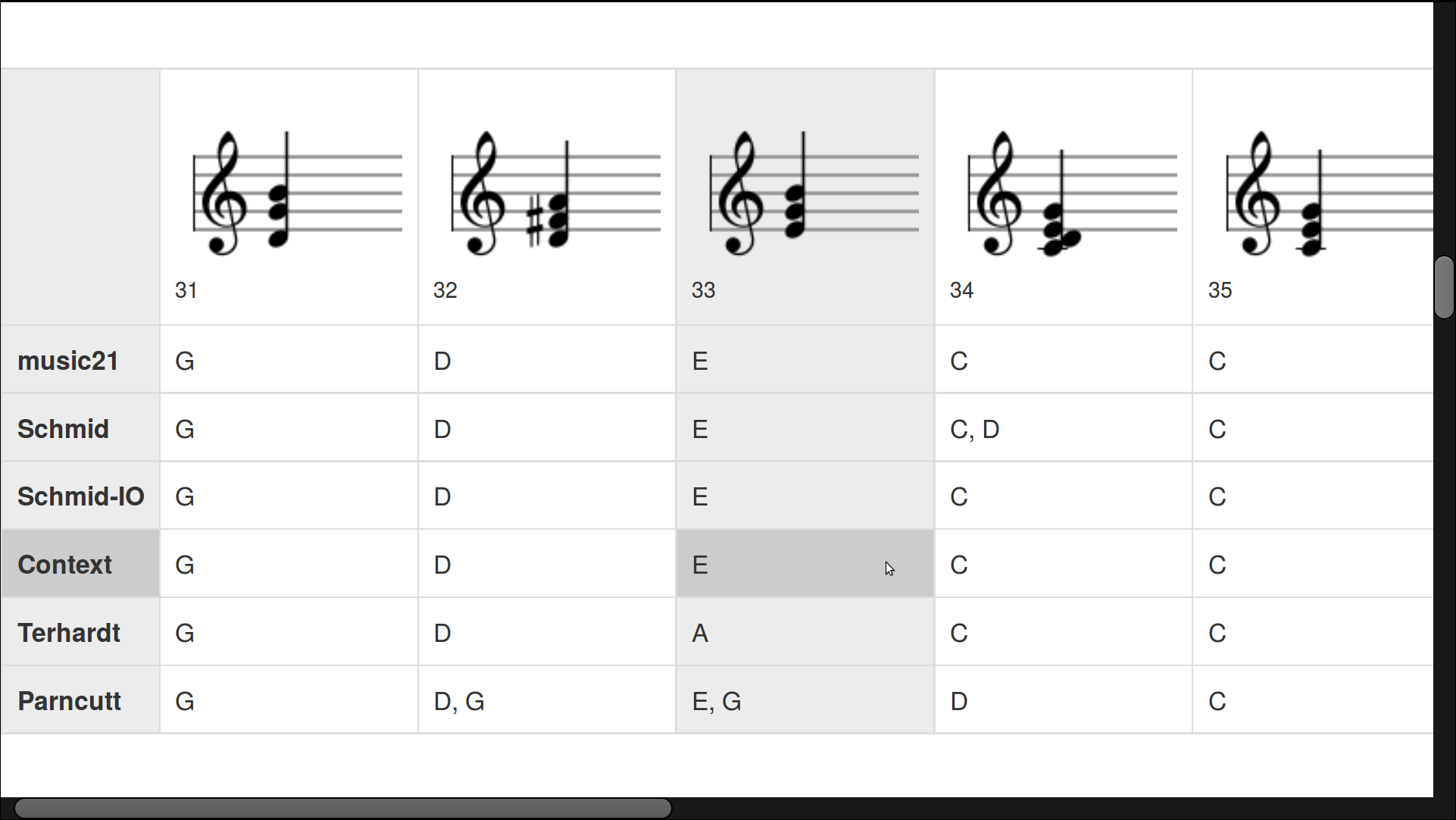}
}
    \end{center}
\label{fig:HTMLOutputExample}
\caption{A screenshot of the HTML output of parts of BWV 99.6 (in \emph{G major}).}
\begin{center}
\small
Note that the displayed notes only represent the pitch classes of the chords.\\
The bass notes are $G$, $F\sharp$, $E$, $B$ and $C$.
\end{center}
\end{figure}

The GUI is minimalistic and only provides the basic options of selecting which
files to output and which models to use for the analysis. The files that should
be analysed can be dragged and dropped into the top area.
Figure~\ref{fig:GUIExample} shows a screenshot of the running application.
Figure~\ref{fig:HTMLOutputExample} depicts an example of (parts of) the HTML
output.

\medskip
\medskip

\chapter{Conclusion and Future Work}

\section{Conclusion}

In this thesis we first gave a basic introduction to music theory and in
particular chord roots, for which we described several existing techniques on
how to determine them. Then we introduced the newly developed Schmid model and
showed how we extended it by exploiting sequential context to predict
nonharmonic tones from a Computer Science perspective.  Lastly, we implemented
all the models and showed how researchers can use the program to do their own
analyses of musical pieces.

A big hurdle for using machine learning algorithms for automatic feature
extraction and classification is the lack of training data.
If one day we have enough annotated data, it might be possible to use an
automatic feature extraction and classification algorithm to directly determine
chord roots from sequences of chords\footnote{We could for example input
the sequence of chords as matrix of pitch classes and the output class also as
pitch class, so that we directly get the root of the chords without needing
constructed features using ELF~\citep{Utgoff1996}.}, but for this thesis we had
to do a lot more manual work:
First we searched for possible features of chord pairs and then we built a
decision tree by hand, which works moderately well for pieces from Bach, but
will probably fail for pieces from many different musical epochs.

\section{Future Work}

A small step has been done towards a better understanding of chord roots, but we
are still very far away from a unifying model that works for all kinds of music.
(Which would be more preferable than developing different theories for different
kinds of music.)

While working on this thesis we came across several questions, which can not yet
be answered conclusively and need addressing (Schmid, 2015, personal
conversation):


\begin{itemize}

    \item How do the new models hold up in empirical studies using both tonal
        and atonal music from different musical epochs?

    \item In what way do humans perceive chord roots and in what role do they
        play in the perception of music? Are they primarily a music theoretical
        concept?

    \item Until what maximum time interval does the human ear perceive tones
        that sound immediately after each other as chord? Is the perception of
        chord roots different when the notes do not sound at the same time?

    \item How important is sequential context for the determination of chord
        roots? How big can such a context be and how do we identify its
        boundaries?

    \item Is it possible to always correctly predict a single root when a chord
        is embedded in some sequential harmonic context or can it still be
        ambiguous?

    \item Are absolute pitches important? Are the heights of pitches relative to
        each other important?
        (c.f.~Figure~\ref{fig:FutureWorkExample})

    \item In what way do different musical parameters (tone color, rhythm,
        dynamic, \ldots) influence the perception of chord roots?

\end{itemize}

\bibliography{refs}
\appendix
\chapter{List of Intervals}
\label{chap:Intervals}
\vspace{-0.5cm}
Table~\ref{tab:IntervalsComplete} shows the names of different intervals. $N$
shows the number of semitones (i.e.~the distance between two notes on the piano
keyboard counting black as well as white keys) and $M$ shows the number of
diatonic steps (i.e.~the distance between two notes on the piano keyboard only
counting the white keys, or the number of name changes between the notes
ignoring accidentals). In the left table the intervals are sorted by diatonic
steps and in the right table they are sorted by semitones.

\begin{table}[h]
\begin{center}
\begin{tabular}{|c c l|}
    \hline
    $M$ & $N$ & Interval \\
    \hline
    \multirow{2}{*}{0}
    & 0   & Perfect unison      \\
    & 1   & Augmented unison    \\
    \hline
    \multirow{4}{*}{1}
    & 0   & Diminished second   \\
    & 1   & Minor second        \\
    & 2   & Major second        \\
    & 3   & Augmented second    \\
    \hline
    \multirow{4}{*}{2}
    & 2   & Diminished third    \\
    & 3   & Minor third         \\
    & 4   & Major third         \\
    & 5   & Augmented third     \\
    \hline
    \multirow{3}{*}{3}
    & 4   & Diminished fourth   \\
    & 5   & Perfect fourth      \\
    & 6   & Augmented fourth    \\
    \hline
    \multirow{3}{*}{4}
    & 6   & Diminished fifth    \\
    & 7   & Perfect fifth       \\
    & 8   & Augmented fifth     \\
    \hline
    \multirow{4}{*}{5}
    & 7   & Diminished sixth    \\
    & 8   & Minor sixth         \\
    & 9   & Major sixth         \\
    & 10  & Augmented sixth     \\
    \hline
    \multirow{4}{*}{6}
    & 9   & Diminished seventh  \\
    & 10  & Minor seventh       \\
    & 11  & Major seventh       \\
    & 12  & Augmented seventh   \\
    \hline
    \multirow{2}{*}{7}
    & 11  & Diminished octave   \\
    & 12  & Perfect octave      \\
    \hline
\end{tabular}
\quad \quad \quad
\begin{tabular}{|c c l|}
    \hline
    $N$ & $M$ & Interval \\
    \hline
    \multirow{2}{*}{0}
    & 0   & Perfect unison      \\
    & 1   & Diminished second   \\
    \hline
    \multirow{2}{*}{1}
    & 1   & Minor second        \\
    & 0   & Augmented unison    \\
    \hline
    \multirow{2}{*}{2}
    & 1   & Major second        \\
    & 2   & Diminished third    \\
    \hline
    \multirow{2}{*}{3}
    & 2   & Minor third         \\
    & 1   & Augmented second    \\
    \hline
    \multirow{2}{*}{4}
    & 3   & Major third         \\
    & 4   & Diminished fourth   \\
    \hline
    \multirow{2}{*}{5}
    & 3   & Perfect fourth      \\
    & 2   & Augmented third     \\
    \hline
    \multirow{2}{*}{6}
    & 4   & Diminished fifth    \\
    & 3   & Augmented fourth    \\
    \hline
    \multirow{2}{*}{7}
    & 4   & Perfect fifth       \\
    & 5   & Diminished sixth    \\
    \hline
    \multirow{2}{*}{8}
    & 5   & Minor sixth         \\
    & 4   & Augmented fifth     \\
    \hline
    \multirow{2}{*}{9}
    & 5   & Major sixth         \\
    & 6   & Diminished seventh  \\
    \hline
    \multirow{2}{*}{10}
    & 6   & Minor seventh       \\
    & 5   & Augmented sixth     \\
    \hline
    \multirow{2}{*}{11}
    & 6   & Major seventh       \\
    & 7   & Diminished octave   \\
    \hline
    \multirow{2}{*}{12}
    & 7   & Perfect octave      \\
    & 6   & Augmented seventh   \\
    \hline
\end{tabular}
\caption{Interval names and their corresponding diatonic steps and semitones.}
\label{tab:IntervalsComplete}
\end{center}
\end{table}

\chapter{Benchmarks}
\label{chap:AppendixBenchmarks}


\begin{table}[h!]
\begin{center}

\begin{tabular}{|l|c|c|c|c|c|c|c|}
    \hline
    \textbf{Piece} & \textbf{ST} & \textbf{Schmid} &
    \textbf{IO} & \textbf{Context} & \textbf{Auto} & \textbf{Terhardt} &
    \textbf{Parncutt} \\
    \hline
    BWV 5.7 & 86.44 & 83.05 & 84.75 & 98.31 & 83.05 & 42.37 & 61.02 \\
    BWV 6.6 & 83.33 & 79.63 & 81.48 & 96.30 & 87.04 & 42.59 & 68.52 \\
    BWV 10.7 & 97.92 & 93.75 & 93.75 & 97.92 & 93.75 & 62.50 & 56.25 \\
    BWV 13.6 & 86.89 & 80.33 & 81.97 & 98.36 & 96.72 & 59.02 & 55.74 \\
    BWV 14.5 & 83.33 & 80.00 & 80.00 & 100.00 & 85.56 & 41.11 & 60.00 \\
    BWV 17.7 & 88.98 & 86.51 & 86.51 & 96.03 & 88.10 & 58.73 & 54.76 \\
    BWV 18.5 & 81.31 & 71.03 & 71.96 & 93.46 & 74.77 & 30.84 & 57.94 \\
    BWV 20.7 & 86.02 & 77.42 & 77.42 & 93.55 & 80.65 & 43.01 & 51.61 \\
    BWV 26.6 & 82.26 & 82.26 & 82.26 & 91.94 & 83.87 & 45.16 & 72.58 \\
    BWV 32.6 & 80.21 & 72.92 & 73.96 & 100.00 & 81.25 & 59.38 & 59.38 \\
    BWV 38.6 & 87.84 & 86.49 & 87.84 & 98.65 & 75.68 & 36.49 & 63.51 \\
    BWV 44.7 & 85.00 & 83.33 & 83.33 & 98.33 & 98.33 & 50.00 & 43.33 \\
    BWV 55.5 & 84.95 & 76.34 & 79.57 & 100.00 & 84.95 & 44.09 & 55.91 \\
    BWV 99.6 & 85.14 & 75.68 & 81.08 & 97.30 & 87.84 & 56.76 & 54.05 \\
    BWV 120.6 & 84.78 & 79.35 & 81.52 & 97.83 & 84.78 & 58.70 & 59.78 \\
    BWV 139.6 & 81.94 & 73.61 & 76.39 & 98.61 & 87.50 & 48.61 & 51.17 \\
    BWV 151.5 & 86.54 & 75.00 & 78.85 & 90.38 & 86.54 & 55.77 & 50.00 \\
    BWV 190.7 & 96.52 & 96.52 & 96.52 & 100.00 & 98.26 & 64.35 & 50.43 \\
    BWV 244.15 & 89.74 & 78.21 & 80.77 & 93.59 & 80.77 & 53.85 & 53.85 \\
    BWV 244.54 & 85.53 & 76.32 & 76.32 & 94.74 & 90.79 & 44.74 & 63.16 \\
    BWV 310 & 86.00 & 80.00 & 82.00 & 96.00 & 86.00 & 44.00 & 72.00 \\
    BWV 312 & 77.05 & 76.23 & 76.23 & 90.16 & 80.33 & 33.61 & 67.21 \\
    BWV 418 & 94.12 & 90.59 & 91.76 & 95.29 & 88.24 & 52.94 & 55.29 \\
    BWV 419 & 87.36 & 77.01 & 78.16 & 87.36 & 72.41 & 29.89 & 62.07 \\
    BWV 429 & 86.21 & 79.31 & 82.76 & 90.80 & 86.21 & 57.47 & 54.02 \\
    BWV 438 & 89.80 & 73.47 & 79.59 & 79.59 & 73.47 & 55.10 & 65.31 \\
    \hline
    \textbf{Overall} & \textbf{86.13} & \textbf{80.21} & \textbf{81.67} &
    \textbf{95.34} & \textbf{85.01} & \textbf{48.59} & \textbf{58.20} \\
    \hline
\end{tabular}

\end{center}
\label{tab:Benchmark}
    \caption{Achieved correctness of the different corpus on the annotated Bach
    corpus in percent.}
    \begin{center}
    \small
    ``ST'' is stacking thirds, ``IO'' is the Schmid model
    with an interval order and ``Auto'' is the Context model with the
    automatically generated decision tree. The annotated corpus contains
    question marks for some chords of which the correct roots were unknown.
    These chords are ignored for this benchmark.
\end{center}
\end{table}

\chapter{List of Program Options}
\label{chap:AppendixProgramOptions}

To get a list of all possible options, we can use the ``help'' flag,
which will give us a detailed description.

\begin{lstlisting}[basicstyle=\ttfamily]
  $ python3 analyse.py -h
\end{lstlisting}

\noindent The general usage of the program is

\begin{lstlisting}[basicstyle=\ttfamily]
  $ python3 analyse.py INPUT [OPTIONS]
\end{lstlisting}

\begin{table}[h!]
\begin{center}

\begin{tabular}{|l|l|}
    \hline
    Option (Short Option) & Description \\
    \hline
    \verb|--help (-h)| & Show a help message and exit \\
    \verb|--filetype (-t) FILETYPE| & Set the filetype of the files that will be read
    in from the input-\\&directory (default is \verb|.mxl|)\\
    \verb|--outdir (-o) OUTDIR| & Choose the output-directory (By default all
    outputfiles will be\\& placed in the same directory as the input-files) \\
    \verb|--models (-m) MODELS| & Choose the models with which the pieces will be
    analysed.\\& \verb|MODELS| should be a string containing the model-names\\& seperated by
    whitespaces.\\& E.g.: \verb|'music21 Parncutt Terhardt Schmid Context'|\\
    \verb|--nohtmls| & Do not output \verb"HTML" file of the analysis \\
    \verb|--musicxmls (-mx)| & Output a numbered \verb|MusicXML| file for each inputfile \\
    \verb|--txts| & For each inputfile write out \verb".txt" files for each models
    contain-\\&ing the predicted roots (one root per line) \\

    \verb|--statistics (-s)| & Calculate and display statistics of correctly
    predicted roots for \\&each model. This needs a \verb".txt" file for each inputfile
    where the \\&correct roots are contained (one root per line). The name of
    \\& the file should be \\&\verb|name_of_inputfile_without_suffix + .correct.txt|\\
    \verb|--debug (-d)| & Display debug messages \\
    \verb|--verbose (-v)| & Increase the output verbosity \\
    \hline
\end{tabular}

\end{center}
    \label{tab:ProgramOptions}
    \caption{List of possible program options.}
\end{table}

\end{document}